\documentclass[final,3p,times,onecolumn,number,review,reviewsort&compress]{elsarticle}
\usepackage{amssymb}
\usepackage{upgreek}
\usepackage{hyperref}
\usepackage{makecell}

\journal{Nuclear Instruments and Methods: A}

\begin{document}

\begin{frontmatter}

\title{A plastic scintillator array for reactor based anti-neutrino studies}

\author[a,b]{D.~Mulmule\corref{cor1}}
\ead{dhruvm@barc.gov.in}
\cortext[cor1]{Corresponding author}
\author[a]{S.~P.~Behera}
\author[a]{P.~K.~Netrakanti}
\author[a]{D.~K.~Mishra}
\author[c]{V.~K.~S.~Kashyap}
\author[a]{V.~Jha}
\author[a,b]{L.~M.~Pant}
\author[a,b]{B.~K.~Nayak}
\author[a,b]{A.~Saxena}

\address[a]{Nuclear Physics Division, Bhabha Atomic Research Centre, Trombay, Mumbai - 400085}
\address[b]{Homi Bhabha National Institute, Anushakti Nagar, Mumbai - 400094}
\address[c]{School of Physical Sciences, National Institute of Science Education and Research, HBNI, Jatni,Khurda - 752050}
\begin{abstract}
  Indian Scintillator Matrix for Reactor Anti-Neutrinos (ISMRAN), a plastic scintillator array (10$\times$10), is being constructed for the purpose of electron anti-neutrino ($\overline\nuup_{e}$) detection for reactor monitoring applications. A prototype detector called mini-ISMRAN, which consists of 16$\%$ of ISMRAN, has been setup for studying the detector response, background rates and event characterization in the reactor and non-reactor environment. The data acquisition system based on waveform digitizers is being used for pulse processing and event triggering. Monte-Carlo based simulations using GEANT4 are performed to optimize lead (Pb) and borated polyethylene (BP) shielding for background reduction and to study the positron, neutron and $\gamma$-ray response in the ISMRAN detector. Characterization of plastic scintillator detectors with known radioactive sources is performed for energy, timing and position measurements. Using the energy summation and bar multiplicity selection, coincident events from $\mathrm{{}^{60}Co}$ decay are reconstructed in non-reactor environment. Results from background measurements using various detectors are quantified in reactor ON and OFF condition. The shielding of 10 cm Pb and 10 cm BP along with the requirement of hits in multiple bars, reduces the uncorrelated background in reactor ON condition. 
\end{abstract}

\begin{keyword}


  Reactor monitoring \sep Anti-neutrino \sep Plastic scintillators.
\end{keyword}

\end{frontmatter}


\section{Introduction}
Nuclear reactors are the most intense source of anti-neutrinos available today. The electron anti-neutrino ($\overline\nuup_{e}$) flux from a 1.0 $\mathrm{GW}$ thermal power output reactor is approximately $\mathrm{10^{21}}$ $\overline\nuup_{e}/s$. Anti-neutrinos are highly penetrating as their interaction cross section with matter is extremely small. The possibility of monitoring nuclear reactors non-intrusively from a remote location, using the weakly interacting nature of neutrinos and the high reactor flux has been demonstrated before~\cite{VVER}. However, the feasibility of such an exercise with relatively small detectors, $\mathrm{\sim}$1.0 ton liquid scintillator (LS), designed for long-term  unmanned operation was established only recently~\cite{SONGS}. These detectors have been shown to be sensitive to the reactor ON and OFF cycles. Also, since the $\overline\nuup_{e}$ rate and energy spectrum changes as the uranium in the core is consumed and plutonium is produced, it is possible to calculate the burn up and estimate the isotopic content of the core~\cite{ADD}. Several groups across different countries are already pursuing this activity~\cite{KURODA,NUCIFER,ANGRA}. This technique of monitoring reactors remotely may be useful for the International Atomic Energy Agency's (IAEA) `Reactor Safeguards Regime' aimed towards ensuring implementation of safeguards for reactor facilities~\cite{iaea}.

Apart from relatively small volume of the detector, factors such as mobility of the setup, safety and convenience of use, especially, from the point of view of long-term operation are crucial for the goal of reactor monitoring. Due to their chemical composition, LS are toxic, flammable and face issues of compatibility with the container material, as they are good solvents. Plastic scintillators (PS) on the other hand are 98$\%$ Polyvinyl chloride (PVC), Polyvinyl Toluene (PVT) or polystyrene i.e. similar to normal plastic with no toxic or radioactive component and non-flammable. Therefore, for long-term near reactor operation use of PS is preferable. However, PS suffers from issues such as reduced light output due to attenuation and radiation damage. These aspects have been extensively studied and addressed to a reasonable extent in modern commercially available PS detectors~\cite{knoll}. Also, majority of PS can not use pulse shape discrimination (PSD) technique for discrimination between neutron and $\gamma$-ray signals. However, a segmented geometry of many plastic scintillator bars, employing a thermal neutron capture agent, can make use of the hit patterns and energy deposition profile to separate neutron capture events from background~\cite{OGURI}.

In addition to the monitoring possibilities, reactor experiments are also good probes for fundamental neutrino physics such as precise measurement of neutrino oscillation parameters~\cite{kamLAND,DChooz,DayaBay,RENO} or search for existence of sterile neutrinos. With improved calculations available for the reactor $\overline\nuup_{e}$ spectrum one can confirm a deficit between their measured and expected flux, referred to as Reactor Anti-Neutrino Anomaly (RAA)~\cite{Mueller,huberspect,Mention}, has prompted a search for oscillation of these neutrinos into sterile states at short baselines~\cite{NEOS,DANSS,PROSPECT}. Moreover, majority of the above reactor experiments have observed a  distinct bump in the $\overline\nuup_{e}$ spectrum at $\sim$5 MeV energy re-invoking a fresh review of the fission and beta decay processes in the reactor core~\cite{Langford,IGISOL,Hayes,Sonzogni,Rasco,HuberFive}.

In view of the above motivations, a feasibility study has been performed which involved simulations to determine the type and configurations of the detectors to be utilized for realizing the above physics~\cite{varchaswi}. From the conclusions drawn as part of the study, a mobile, plastic scintillator array - `ISMRAN (Indian Scintillator Matrix for Reactor Anti-Neutrinos)' with a total weight of 1.0 ton is being constructed, for purpose of reactor monitoring and sterile neutrino searches at the Dhruva research reactor facility in Bhabha Atomic Research Centre (BARC), Trombay, India. 

In this paper we discuss the details of the full ISMRAN setup and its geometrical configuration. We present the characterization studies, response and background measurements pertaining to the basic array element of ISMRAN, a PS bar, in reactor and non-reactor environment. Results from a prototype detector, mini-ISMRAN, which is a smaller matrix of 16 PS bars i.e. 16$\%$ of the full ISMRAN are presented. GEANT4 simulations performed to determine event selection criteria and for evaluating effectiveness of the proposed shielding are also presented.

\section{The ISMRAN Array}
\begin{figure}[h]
\begin{center}
\includegraphics[scale=0.40]{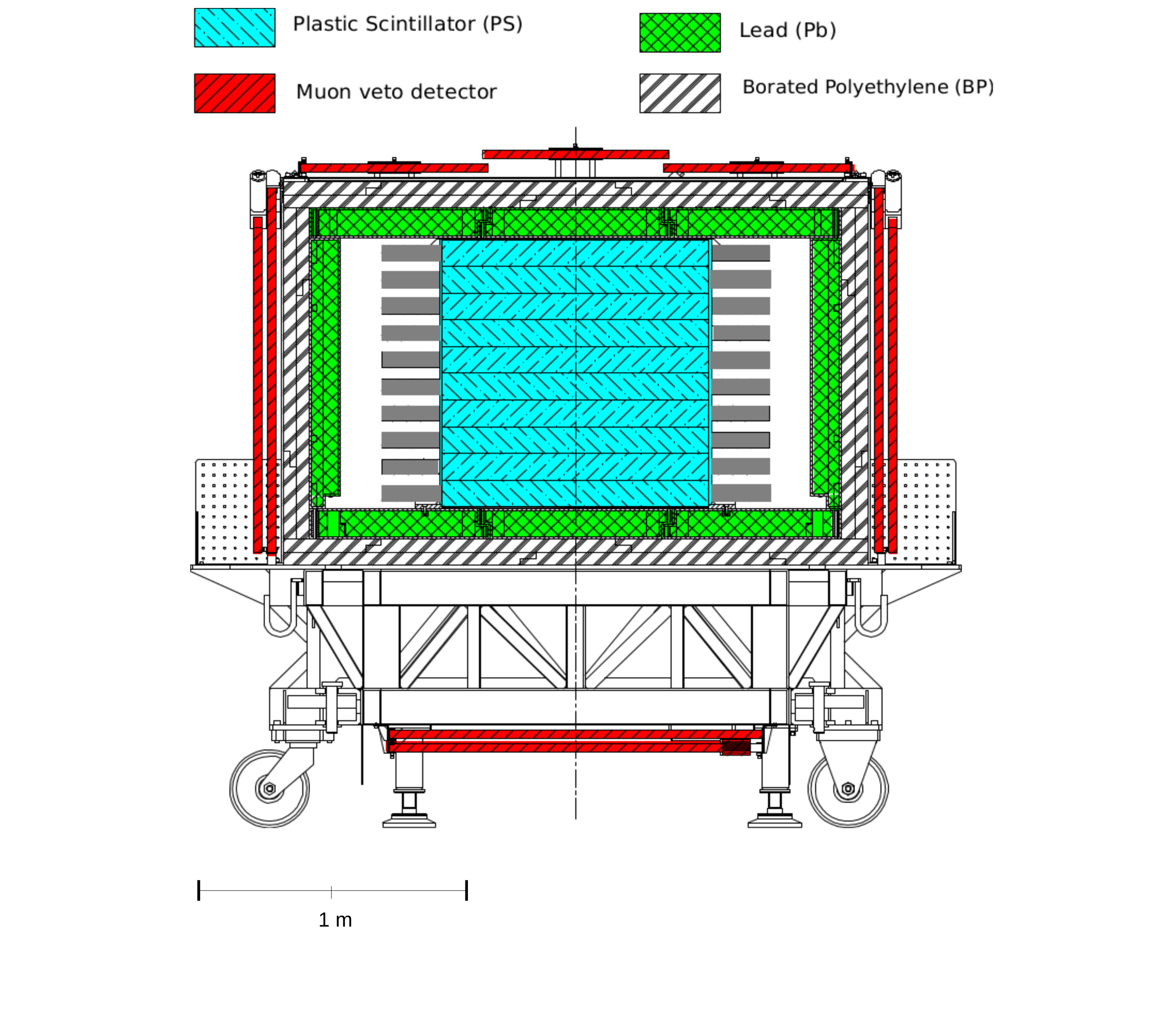}
\caption{Proposed ISMRAN detector setup comprising of shielding trolley and 100 PS bars. The major components of the setup are listed in their respective colors at top.}
\label{ISMRAN100}
\end{center}
\end{figure}

ISMRAN array is an above ground plastic scintillator (PS) based detector of weight 1.0 ton. It comprises of 100 PS bars of dimension $\mathrm{100 cm \times 10 cm \times 10 cm}$ as shown in  Fig.\ref{ISMRAN100}. Each PS bar is coupled to a 3'' PMT at both ends and wrapped with Gadolinium oxide ($\mathrm{Gd_{2}O_{3}}$) coated aluminized mylar foils. The PS bars are EJ200 and PMTs used are ETL 9821 series tubes. The areal density of $\mathrm{Gd_{2}O_{3}}$ is $\mathrm{4.8 mg/cm^{2}}$ which corresponds to $\sim$0.18$\%$ of the total detector weight. The location of the setup is at a distance of $\sim$13m from Dhruva reactor core with a flexibility of movement using the trolley stand structure.
The Dhruva research reactor facility at Bhabha Atomic Research Centre (BARC) houses a thermal reactor ($\mathrm{100 MW_{th}}$) with a natural uranium core in the form of a cluster of fuel rods. It is a vertical tank type reactor with separate channels of heavy water ($\mathrm{D_{2}O}$) used as moderator and coolant. It is designed to provide a high neutron flux $\sim$$\mathrm{10^{14}~n/cm^{2}/s}$ (at the core center), for use in radio-isotope production, neutron activation analysis and testing of neutron detectors.
While operating the ISMRAN setup above ground and close to Dhruva reactor core, the detected $\overline\nuup_{e}$ signal suffers from natural as well as reactor related background sources. A passive shielding of 10 cm lead (Pb) inside followed by 10 cm of borated polyethylene (BP), with 30$\%$ boron concentration, outside is being incorporated to reduce these backgrounds. Furthermore, to reduce the background from cosmic muon events an outer layer consisting of 3 cm thick plastic scintillators are used as muon veto detectors.
\section{Detection principle and expected anti-neutrino rate}
Reactor anti-neutrinos are primarily detected via the inverse beta decay (IBD) process which has an interaction cross-section of approximately $\mathrm{6 \times 10^{-43}cm^{2}}$. In this reaction an $\overline\nuup_{e}$ interacts with a proton to produce a positron and a neutron as follows:
\begin{equation}
\mathrm{ \overline \nuup_{e} + p \rightarrow e^{+} + n}.
\end{equation}
The above process has an energy threshold of 1.806 MeV. The positron, which carries almost all of the energy of the neutrino, loses energy in the scintillator volume through ionization and annihilation with an electron and produces two $\gamma$-rays of 0.511 MeV. The signal produced by the positron and the resulting $\gamma$-rays provide the prompt event.
On the other hand, the neutron which has only a few keVs of energy undergoes thermalization in the detector volume and finally may get captured on either Gd or H nuclei. These thermal neutron captures are more likely to occur on the wrapped Gd foil on the sides of PS bar due to the very high thermal neutron capture cross-section ($\sigmaup_{n-capture}$) of Gd as compared to that of H.
\begin{equation}\label{eq:Gd1}
  \hspace{-0.2in}
\mathrm{ n + {}^{155}Gd \rightarrow {}^{156}Gd^{*} \rightarrow \gamma 's,   \quad \sum E_{\gamma} = 8.5~MeV}, \quad \sigmaup_{n-capture} = \mathrm{61000~b},
\end{equation}
\begin{equation}\label{eq:Gd2}
  \hspace{-0.2in}
\mathrm{ n + {}^{157}Gd \rightarrow {}^{158}Gd^{*} \rightarrow \gamma 's,  \quad \sum E_{\gamma} = 7.9~MeV}, \quad \sigmaup_{n-capture} = \mathrm{254000~b}.
\end{equation}
Neutron capture on Gd leads to a cascade of $\gamma$-rays of total energy $\sim$8 MeV as shown in equations ~\ref{eq:Gd1} and ~\ref{eq:Gd2} . In case, the neutron gets captured on a hydrogen nuclei it will produce a mono-energetic $\gamma$-ray of 2.2 MeV. The neutron capture followed by emission of $\gamma$-ray(s) is more likely to happen after the neutron has thermalized to lower energies and is defined as a delayed event. Electron anti-neutrino events in the detector volume are, therefore, characterized by detection of these prompt and delayed event pairs. The mean time delay between the positron and neutron events can range from ten to hundreds of microsecond depending on the capture agent, its position, and concentration in the scintillator volume.

Electron anti-neutrinos are emitted in the decay chains of fission products of Uranium (U) and Plutonium (Pu) accompanied by release of energy in the reactor. The nuclei ${}^{235}\mathrm{U}$, ${}^{239}\mathrm{Pu}$, ${}^{238}\mathrm{U}$ and ${}^{241}\mathrm{Pu}$ together contribute 99.9$\%$ of the total thermal power~\cite{HuberIso} with major contribution of the emitted $\overline\nuup_{e}$ coming from the fission of ${}^{235}\mathrm{U}$ and ${}^{239}\mathrm{Pu}$ isotopes. The rate of interaction of reactor $\overline\nuup_{e}$ with energies $\mathrm{E_{\overline\nuup_{e}}}$ inside the scintillator volume depends on the $\overline\nuup_{e}$  spectrum per fission $\mathrm{f(E_{\overline\nuup_{e}})}$, the number of free protons $\mathrm{N_{p}}$, detector efficiency $\etaup$, thermal power $\mathrm{P_{th}}$ of the reactor (MW), average energy per fission $\mathrm{\overline E_{f}}$ (MeV) released in the reactor core and the distance $\mathrm{D}$ (cm) between the detector and center of the core.
Hence, the total interaction rate in the detector volume obtained by integrating over energies~\cite{VVER} is given as
\begin{equation}
\mathrm{N_{\overline\nuup_{e}} = \frac{N_{p} \cdot P_{th} \cdot \overline\sigmaup_{IBD} \cdot \etaup} {4 \piup D^{2} \cdot \overline E_{f} \cdot 1.6 \cdot 10^{-19}}},
\end{equation}
where, $\mathrm{\overline\sigmaup_{IBD}=\int\sigmaup(E_{\overline\nuup_{e}})f(E_{\overline\nuup_{e}})d N_{\overline\nuup_{e}}(E_{\overline\nuup_{e}})}$
is the cross section of IBD averaged over the $\overline\nuup_{e}$ spectrum. The cross-section  $\mathrm{\overline\sigmaup_{f}}$ and energy released $\mathrm{\overline E_{f}}$ are usually expressed in terms of the corresponding quantities  $\mathrm{\sigmaup_{i}}$ and $\mathrm{E_{i}}$ for the four dominant isotopes, i.e. ${}^{235}\mathrm{U}$, ${}^{239}\mathrm{Pu}$, ${}^{238}\mathrm{U}$ and ${}^{241}\mathrm{Pu}$, along with  $\mathrm{\alphaup_{i}}$ as the contribution of each isotope to the total number of fissions.
\begin{equation}\mathrm{ \overline\sigmaup_{IBD} = \sum \alphaup_{i}\sigmaup_{i} \\ \overline E_{f} = \sum \alphaup_{i} E_{i}}, \end{equation}
where  $\mathrm{\sum \alphaup_{i}}$=1.
Thus both the thermal power $\mathrm{P_{th}}$ and the contribution from the isotopes to the cross-section ( $\mathrm{\alphaup_{i}\sigmaup_{i}}$) are directly reflected in the $\overline\nuup_{e}$ rate in the detector. For the ISMRAN setup which is at 13 m distance from a 100 $\mathrm{MW_{th}}$ reactor the estimated $\overline\nuup_{e}$ event rate is $\sim$115 per day. This $\overline\nuup_{e}$ event rate is calculated using the fission fractions for a typical CANDU reactor~\cite{Zhan} assuming a compact core and a 30 $\%$ detection efficiency.
\section{Detector Simulation}
\begin{figure}[h]
  \begin{center}
\includegraphics[scale=0.30]{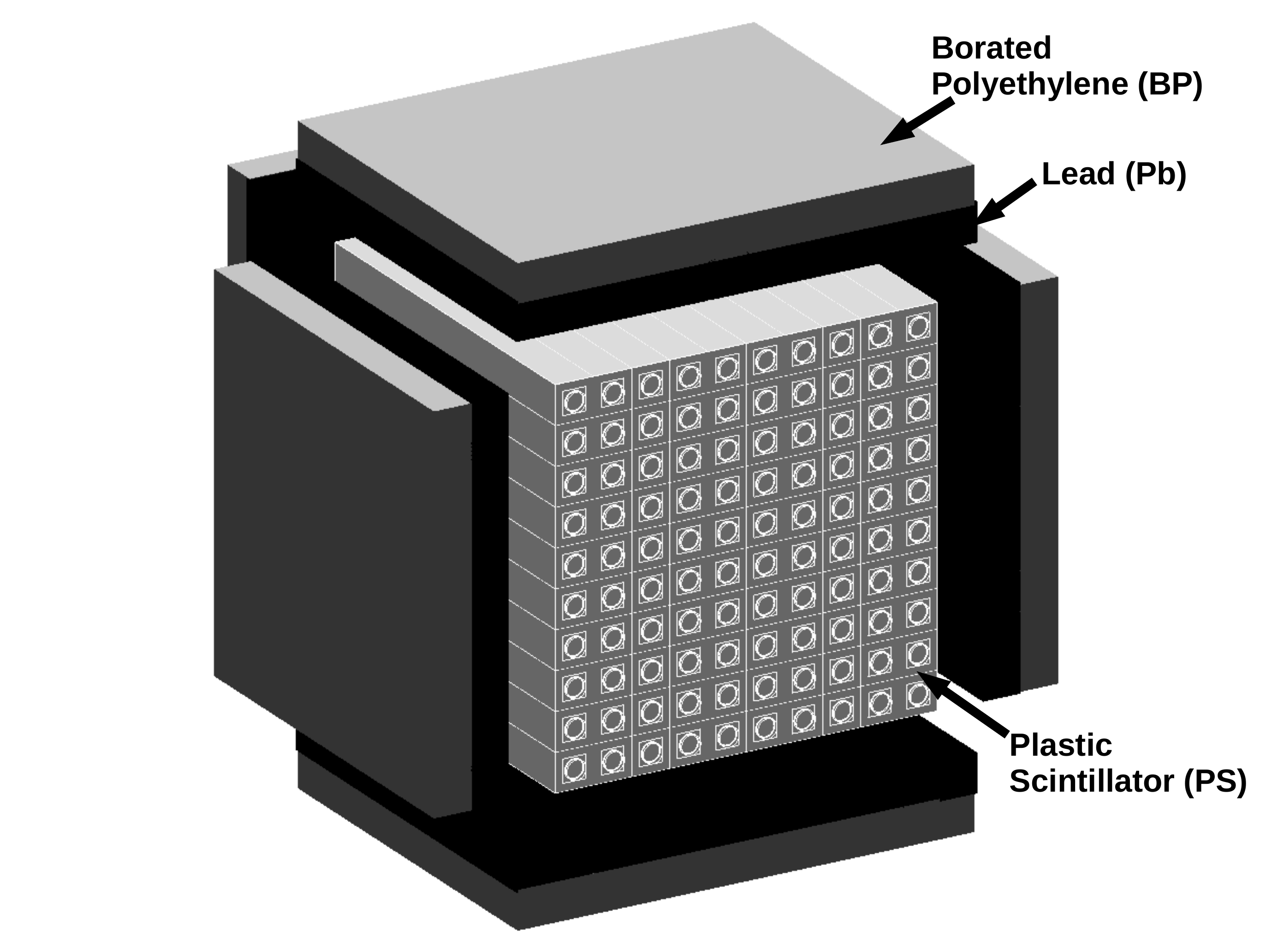}
\caption{ISMRAN 10$\times$10 detector geometry with PS bar setup in GEANT4. Also shown are 10 cm thick Pb bricks and 10 cm thick BP sheet.}
\label{ISMRANDesign}
\end{center}
\end{figure}
\begin{figure*}
\begin{center}
\includegraphics[scale=0.65]{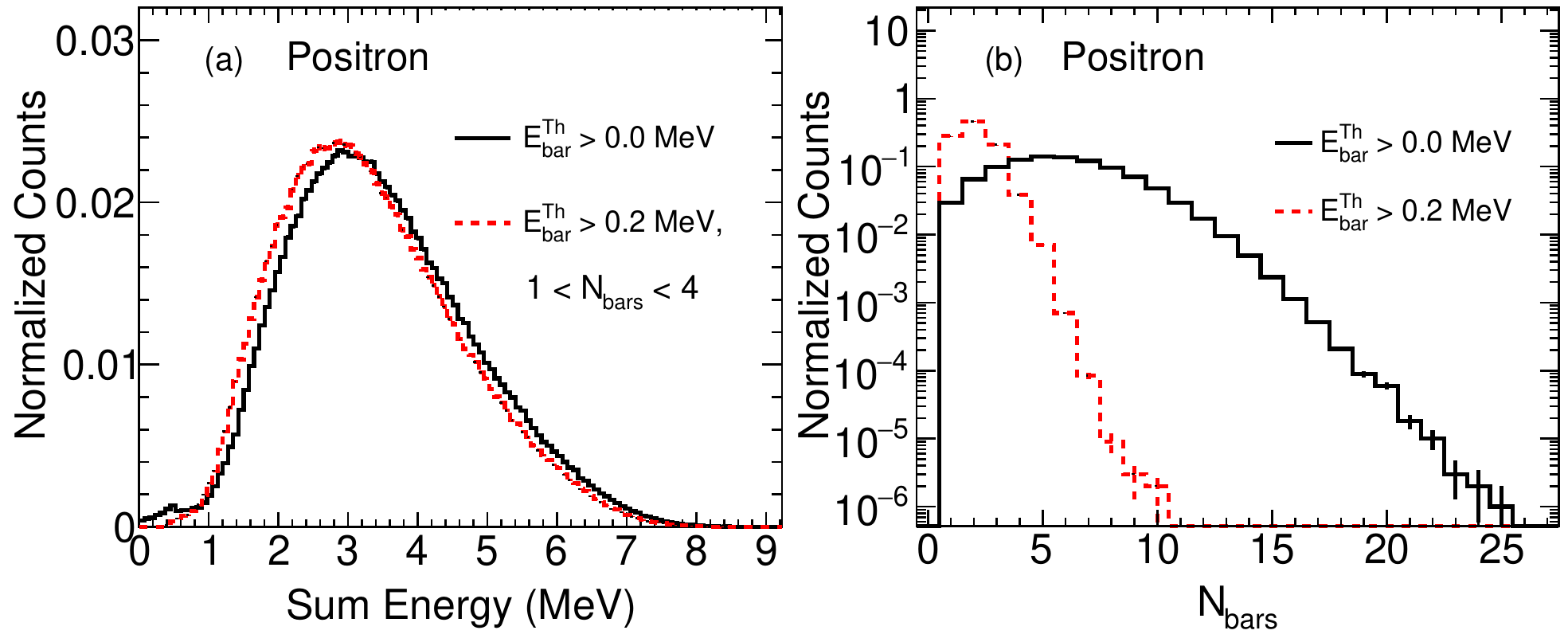}
\caption{(a) The sum energy distribution in PS bars for prompt positron events with $\mathrm{E^{Th}_{bar} > }$  0.0 MeV (solid histogram). Also shown is the effect of $\mathrm{E^{Th}_{bar} > }$  0.2 MeV and number of bars selected ( 1 $\mathrm{< N_{bars} <}$ 4 ) on prompt energy distribution (dashed histogram). Panel (b) shows the $\mathrm{N_{bars}}$ distribution for $\mathrm{E^{Th}_{bar} > }$  0.0 MeV (solid histogram) and $\mathrm{E^{Th}_{bar} > }$  0.2 MeV (dashed histogram).}
\label{PromptPosiSum}
\end{center}
\end{figure*}
\begin{figure*}[t]
\begin{center}
\includegraphics[scale=0.65]{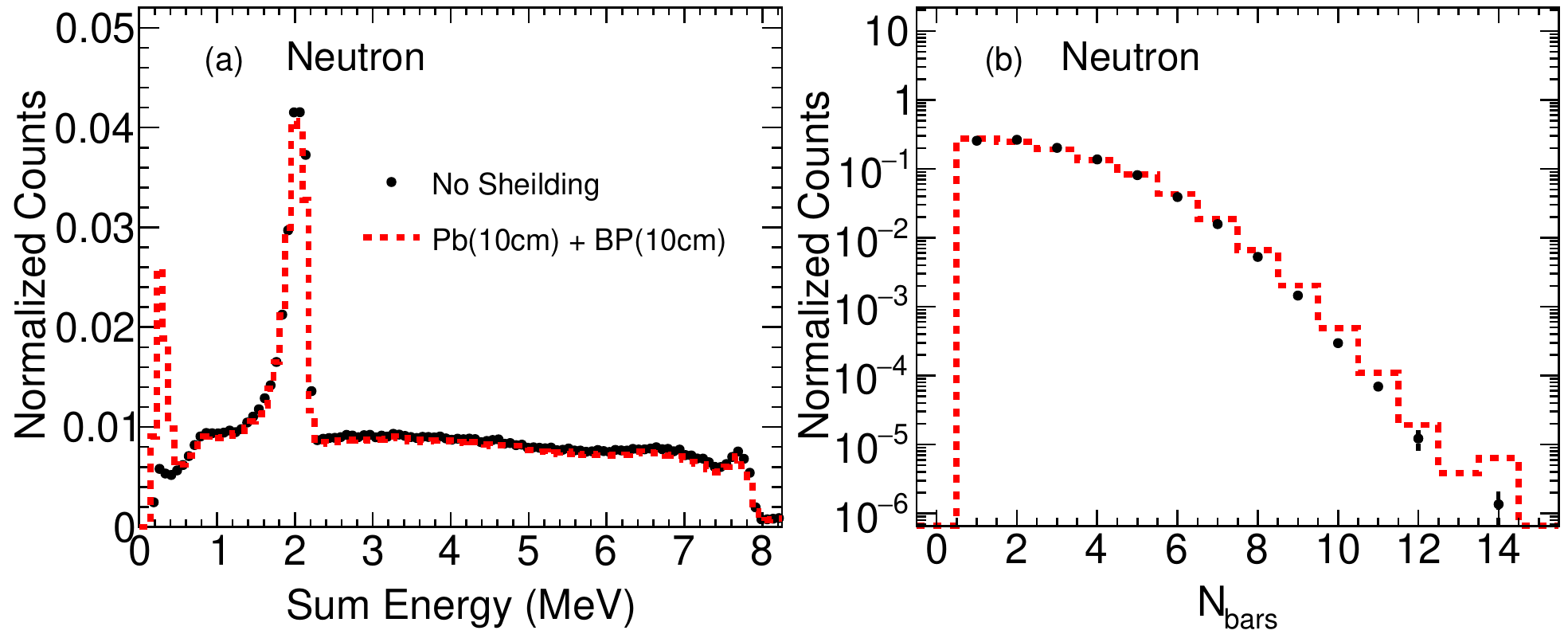}
\caption{(a) The sum energy distribution of delayed neutron events in PS bars with $\mathrm{E^{Th}_{bar} >}$  0.2 MeV for two configurations: no shielding (dotted histogram) and with 10 cm Pb and 10 cm BP shielding (dashed histogram) and (b) the corresponding $\mathrm{N_{bars}}$ hit distribution.}
\label{DelayNeutCapSum}
\end{center}
\end{figure*}

\begin{figure*}
\begin{center}
\includegraphics[scale=0.65]{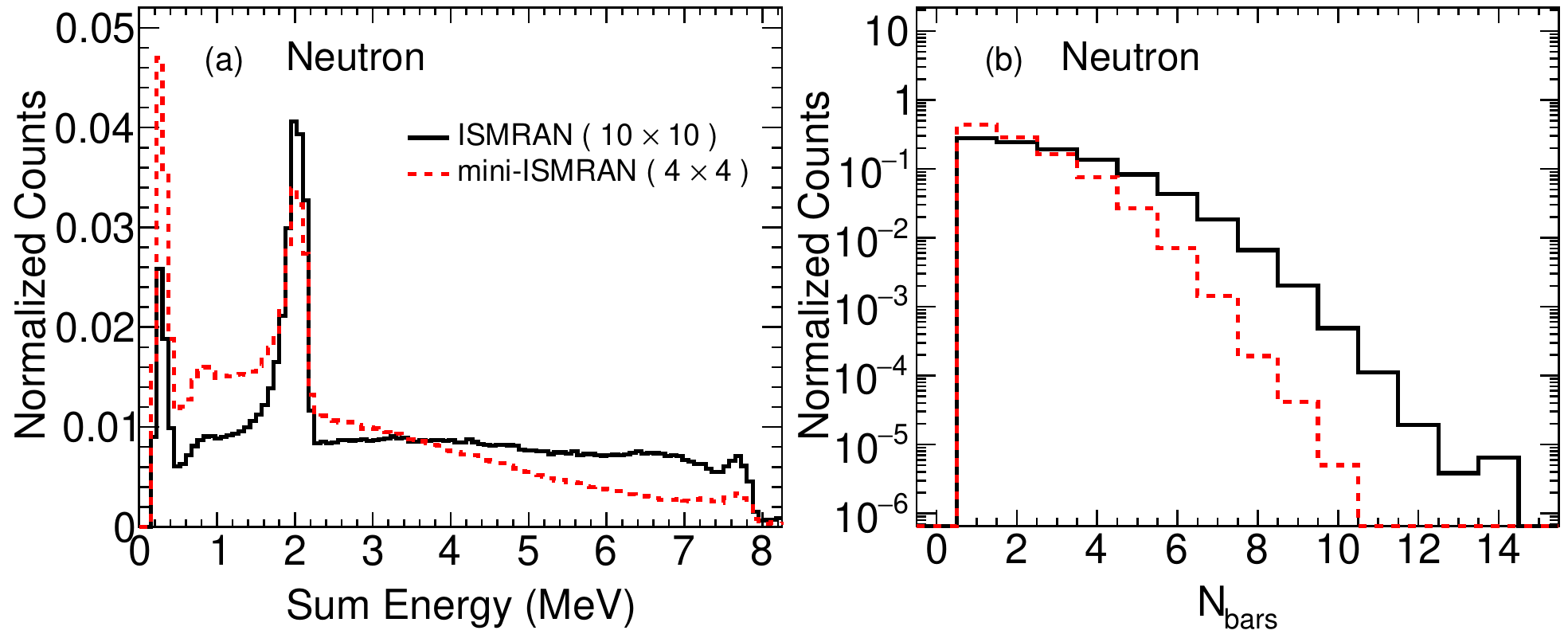}
\caption{(a) Comparison between the sum energy distribution obtained in ISMRAN and miniISMRAN for delayed neutron events and (b) the corresponding $\mathrm{N_{bars}}$ hit distribution in the two geometries.}
\label{ISMRAN_miniISMRAN}
\end{center}
\end{figure*}
Monte-Carlo based simulations using GEANT4~\cite{GEANT4} (version 4.10.3) are carried out to understand the detector response and shielding effects on the detected $\overline\nuup_{e}$  events via IBD in ISMRAN. Figure~\ref{ISMRANDesign} shows the cross-sectional view of 10$\times$10 PS bar detector geometry of ISMRAN setup in GEANT4. The 10 cm lead (Pb) and 10 cm borated polyethylene (BP) shielding around the PS bars is also shown in black and dark gray, respectively. The high precision QGSP\_BIC\_HP physics processes for the low energy neutron interactions and standard electromagnetic processes for $\gamma$-ray, positron and electron are used in simulation. The QGSP\_BIC\_HP list has the G4NeutronHPCapture model which uses the final state method for de-excitation of Gadolinium  nucleus to produce $\gamma$-ray cascade. Though the final state method in GEANT4 agrees well with the individual $\gamma$-ray spectrum from the cascade, it breaks Q-value in low energy events. However, the improvement in the implementation of cascade $\gamma$-rays emission model from de-excitation of Gd nucleus has led to better description of the cascade $\gamma$-ray spectrum~\cite{Yano}.
The energy distribution, number of bars hit ($\mathrm{N_{bars}}$) and timing information for prompt and delayed events are studied for better understanding of the IBD event characteristics in ISMRAN. An energy threshold, $\mathrm{E^{Th}_{bar} > }$ 0.2 MeV, is applied on the deposited energy of each PS bar in simulation, similar to the threshold applied on calibrated energy of each PS bar in reactor environment. This selection of common threshold on calibrated energy of each PS bar, in reactor data, is applied to achieve spectral uniformity among different bars. 
For simulating the IBD events in ISMRAN, the input energy distribution of $\overline\nuup_{e}$ to GEANT4 is obtained by using the parameterization of $\overline\nuup_{e}$ spectrum from reference~\cite{Mueller}, the calculation of cross sections from reference~\cite{Mention,Vogel} and the fission fractions for different isotopes are taken from reference~\cite{Zhan}. The vertex position of the IBD events are generated randomly in the ISMRAN detector volume. The daughter products, positron and neutron energies, are obtained using two-body kinematics for each neutrino energy on event by event basis. These generated positron and neutron are then propagated in GEANT4 for studying the response of PS bars in ISMRAN detector volume. The sum total of energy and number of PS bars hit ($\mathrm{N_{bars}}$) in each event is obtained by addition of deposited energy in each PS bar with $\mathrm{E^{Th}_{bar} > }$ 0.2 MeV.
Figure~\ref{PromptPosiSum}(a) shows the event by event sum energy distribution in ISMRAN for prompt positron event. The prompt sum energy from PS bars comprises of the energy deposited by positron via ionization energy loss and energy deposited by Compton scattered electrons from annihilation $\gamma$-rays. The prompt sum energy closely follows the $\overline\nuup_{e}$ spectrum above the 1.806 MeV threshold as most of the $\overline\nuup_{e}$ energy is carried by positron in IBD event. Figure~\ref{PromptPosiSum}(b) shows the event by event distribution of the $\mathrm{N_{bars}}$ with (dashed histogram) and without (solid histogram) the detector threshold $\mathrm{E_{bar}^{Th}} > $ 0.2 MeV in the prompt positron events. Most of the prompt events deposit energy within 1 $\mathrm{< N_{bars} <}$ 4. The $\mathrm{N_{bars}}$ distribution peaks at $\sim$2 and even extends to 10 bars in some of the events. Also shown in Fig~\ref{PromptPosiSum}(a), the effect of $\mathrm{E_{bar}^{Th}} >$ 0.2 MeV and selection on $\mathrm{N_{bars}}$ ( 1 $\mathrm{< N_{bars} <}$ 4 ) on the prompt energy spectrum. 
Figure~\ref{DelayNeutCapSum}(a) shows the sum energy distributions of $\gamma$-rays, emanating from neutron capture on either H or Gd, in ISMRAN from delayed neutron events. From simulations in ISMRAN geometry, it is observed that $\sim$73$\%$ of the neutrons are captured on Gd and $\sim$25$\%$ are captured on H. A small fraction of neutrons are captured on Carbon ($\sim$0.5$\%$) and Lead shielding ($\sim$1.0$\%$), while $\sim$0.5$\%$ of all the generated neutrons escape from the setup without getting captured.  Neutron capture on Gd will give a cascade of $\gamma$-rays up to energies of 8 MeV while that on H will result in a mono-energetic $\gamma$-ray of energy 2.2 MeV. A distinct peak at $\sim$2.0 MeV from neutron capture on H in the sum energy distribution is observed in GEANT4 indicating the containment of $\gamma$-ray within the 10 $\times$ 10 ISMRAN setup. A low energy peak at $\sim$0.3 MeV in the sum energy distribution, as shown in dashed histogram in Fig~\ref{DelayNeutCapSum}(a), corresponds to the events where capture $\gamma$-rays escape the scintillator volume and undergo Compton scattering from the shielding material (Pb (10cm) + BP(10cm)) and enters back in the sensitive volume of the detector. This feature in sum energy distribution is absent when the simulations are performed without any shielding material around the active detector volume. Due to the $\mathrm{E_{bar}^{Th}} >$ 0.2 MeV as well as the geometrical acceptance of cascade $\gamma$-rays in ISMRAN the 8 MeV of sum energy is distributed over the entire energy range. Figure~\ref{DelayNeutCapSum}(b) shows the event by event distribution of $\mathrm{N_{bars}}$ from the neutron capture events in ISMRAN. 
The geometrical acceptance of the detector which is reflected in the incomplete containment of cascade $\gamma$-rays is also studied in simulation for the mini-ISMRAN ( $\mathrm{4 \times 4}$ ) setup. Figure~\ref{ISMRAN_miniISMRAN}(a) shows the effect of the limited volume of the prototype mini-ISMRAN detector on the delayed neutron sum energy distribution and Fig.~\ref{ISMRAN_miniISMRAN}(b) shows the $\mathrm{N_{bars}}$ distribution of the neutron capture events. The sum energy distribution in mini-ISMRAN is shifted to lower energy, around 1 MeV, as compared to the full ISMRAN setup. This is due to the partial containment of the cascade $\gamma$-rays in the mini-ISMRAN. This effect can also be seen in the $\mathrm{N_{bars}}$ distribution in both the cases. The enhancement of Compton scattered events in the sum energy distribution at $\sim$0.3 MeV for mini-ISMRAN are also increased as compared to the full ISMRAN simulations due to the edge effects for the mini-ISMRAN. To detect $\overline\nuup_{e}$ candidate events, prompt and delayed events are selected according to the sum energy deposited, with $\mathrm{E_{bar}^{Th}} >$ 0.2 MeV, in the detector and  $\mathrm{N_{bars}}$ hits. The associated efficiencies are calculated using GEANT4 simulations for the ISMRAN setup. The $\overline\nuup_{e}$ detection efficiencies are summarized in table~\ref{Table1} for two sets of selection cuts, loose (Selection 1) and tight (Selection 2), on sum energy ranges, $\mathrm{N_{bars}}$ and time difference for prompt and delayed events. In table~\ref{Table1}, the variables  $\mathrm{E^{prompt}}$ and $\mathrm{E^{delayed}}$  denote the sum energy of prompt and  delayed events while the $\mathrm{N_{bars}^{prompt}}$ and $\mathrm{N_{bars}^{delayed}}$ denote the number of bars hit for prompt and delayed events. The selection cut on the time difference ($\mathrm{\Delta T}$) between prompt and delayed events can be used to reduce the cosmogenic neutron induced backgrounds in ISMRAN. Currently, the simulations lack the incorporation of natural background in evaluating the efficiencies and only present the first estimates of $\overline\nuup_{e}$ detection in ISMRAN. Further work on the improvements in $\overline\nuup_{e}$ detection efficiencies are in progress.
\begin{table}[h]
\begin{small}
  \begin{center}
  \caption{Detection efficiencies of $\overline\nuup_{e}$ events in ISMRAN with different prompt and delayed event selections.}
  \label{Table1}
\begin{tabular}{|c|c|c|c|}
\hline
\makecell{Selection 1} & Efficiency ($\%$) & \makecell{Selection 2} & Efficiency ($\%$)\\
\hline
\makecell{1.8 $<$ $\mathrm{E^{prompt}}$(MeV) $<$ 8.0} & 98 & \makecell{2.2 $<$ $\mathrm{E^{prompt}}$ (MeV) $<$ 8.0} & 96 \\
\hline
\makecell{1.8 $<$ $\mathrm{E^{prompt}}$ (MeV) $<$ 8.0, \\ 1 $<$ $\mathrm{N_{bars}^{prompt}}$ $<$ 4}  & 69 & \makecell{2.2 $<$ $\mathrm{E^{prompt}}$ (MeV) $<$ 8.0, \\ 1 $<$ $\mathrm{N_{bars}^{prompt}}$ $<$ 4}  & 67 \\

\hline
\makecell{0.8 $<$ $\mathrm{E^{delayed}}$ (MeV) $<$ 8.0} & 84 & \makecell{3.0 $<$ $\mathrm{E^{delayed}}$ (MeV) $<$ 8.0} & 56 \\
\hline
\makecell{0.8 $<$ $\mathrm{E^{delayed}}$ (MeV) $<$ 8.0, \\ $\mathrm{N_{bars}^{delayed}}$ $>$ 3}  & 29 & \makecell{3.0 $<$ $\mathrm{E^{delayed}}$ (MeV) $<$ 8.0, \\ $\mathrm{N_{bars}^{delayed}}$ $>$ 3}  & 27 \\
\hline
\makecell{1.8 $<$ $\mathrm{E^{prompt}}$ (MeV) $<$ 8.0, \\ 1 $<$ $\mathrm{N_{bars}^{prompt}}$ $<$ 4 \\ 0.8 $<$ $\mathrm{E^{delayed}}$ (MeV) $<$ 8.0, \\ $\mathrm{N_{bars}^{delayed}}$ $>$ 3} & 20 & \makecell{2.2 $<$ $\mathrm{E^{prompt}}$ (MeV) $<$ 8.0, \\ 1 $<$ $\mathrm{N_{bars}^{prompt}}$ $<$ 4 \\3.0 $<$ $\mathrm{E^{delayed}}$ (MeV) $<$ 8.0, \\ $\mathrm{N_{bars}^{delayed}}$ $>$ 3}  & 18 \\
\hline
\makecell{1.8 $<$ $\mathrm{E^{prompt}}$ (MeV) $<$ 8.0, \\ 1 $<$ $\mathrm{N_{bars}^{prompt}}$ $<$ 4 \\ 0.8 $<$ $\mathrm{E^{delayed}}$ (MeV) $<$ 8.0, \\ $\mathrm{N_{bars}^{delayed}}$ $>$ 3 \\ 4.0 $<$ $\mathrm{\Delta T}$ ($\mu$s) $<$ 200.0} & 19 & \makecell{2.2 $<$ $\mathrm{E^{prompt}}$ (MeV) $<$ 8.0, \\ 1 $<$ $\mathrm{N_{bars}^{prompt}}$ $<$ 4 \\3.0 $<$ $\mathrm{E^{delayed}}$ (MeV) $<$ 8.0, \\ $\mathrm{N_{bars}^{delayed}}$ $>$ 3  \\ 8.0 $<$ $\mathrm{\Delta T}$ ($\mu$s) $<$ 200.0} & 16 \\
\hline
\end{tabular}
\end{center}
\end{small}
\end{table}

\begin{figure}[h]
\hspace{-0.3in.}
\begin{center}
\includegraphics[scale=0.40]{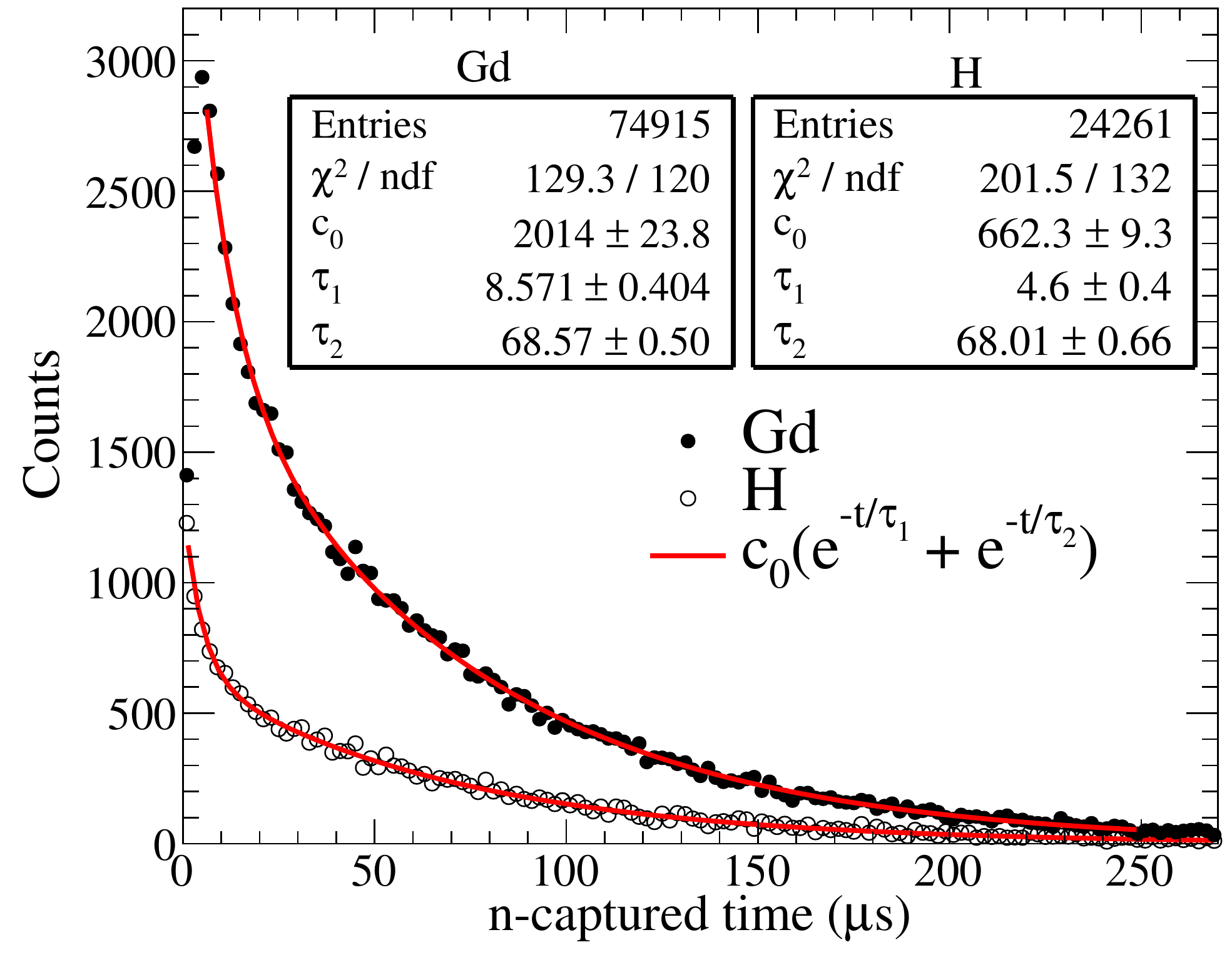}
\caption{$\mathrm{\Delta T}$ distribution between prompt and delayed event in ISMRAN using GEANT4 simulations.}
\label{CapTimeHGd}
\end{center}
\end{figure}

The $\overline\nuup_{e}$ candidate events in ISMRAN can be selected by studying the time correlation between the prompt and delayed events. Using GEANT4, we have studied the capture time of the emitted neutron from IBD events in the ISMRAN detector volume. Figure~\ref{CapTimeHGd}, shows the neutron capture time distribution between the prompt positron annihilation event and $\gamma$-ray(s) from neutron capture either on H or Gd. The neutron capture time distributions are fitted with a double exponential function to obtain the mean neutron capture time of $\sim$68 $\mu$s in ISMRAN setup. The similar mean neutron capture time values on Gd and H in an inhomogenuous detector are discussed in reference~\cite{CaptureNote}.
\begin{figure}[h]
\hspace{-0.1in.}
\begin{center}
\includegraphics[scale=0.38]{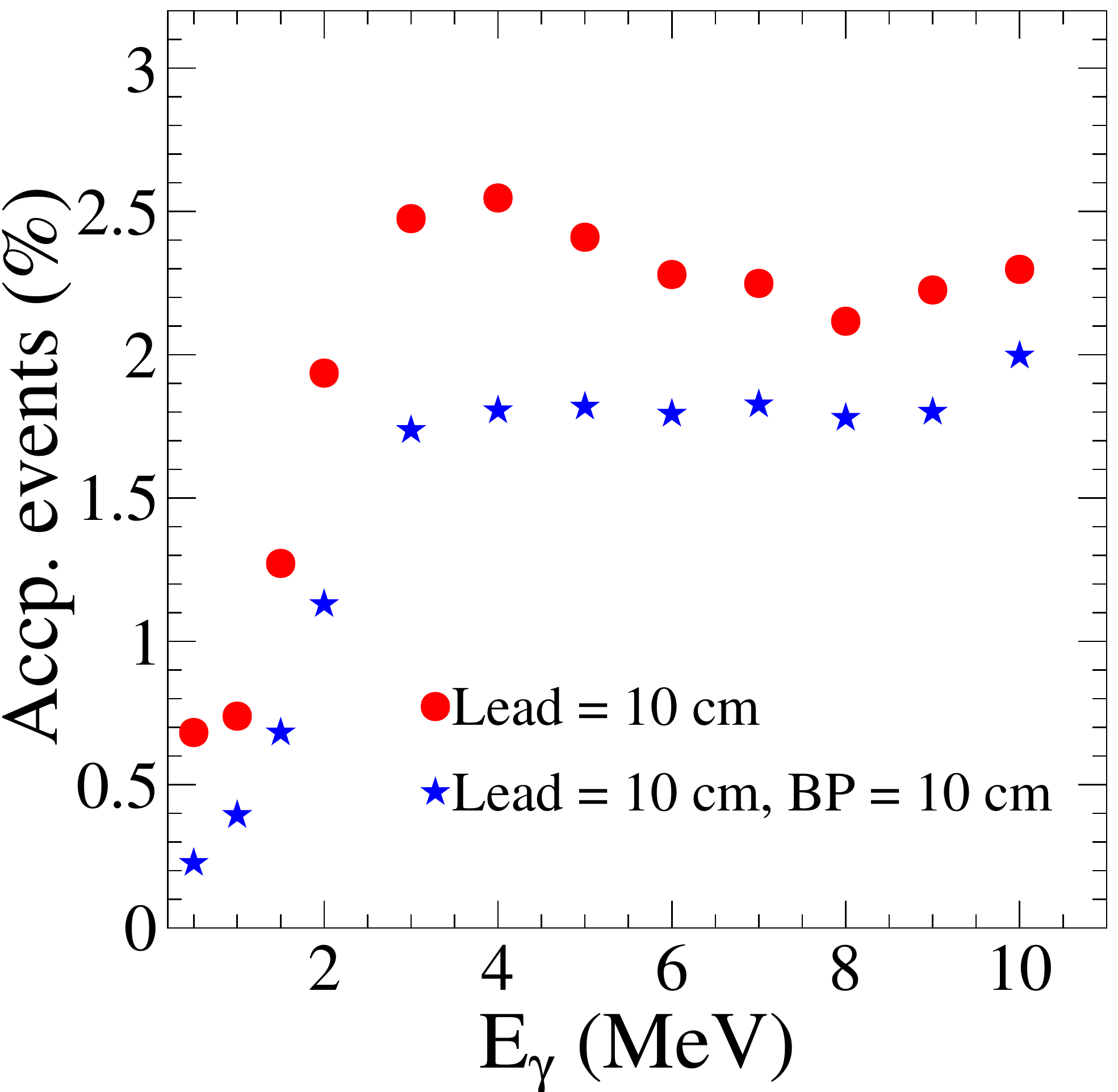}
\caption{Accepted $\gamma$-ray events, inside ISMRAN detector, for two shielding configurations.}
\label{AccepGamma}
\end{center}
\end{figure}

To suppress the ambient and reactor related $\gamma$-ray and neutron background, a study of shielding material consisting of 10 cm of only Pb and Pb in combination with BP (both 10 cm thick) is studied in GEANT4 simulations. The study is performed with $\gamma$-rays, as primary particles in GEANT4, having different energies. Figure~\ref{AccepGamma} shows that incident $\gamma$-energies below 2 MeV have a less than 2$\%$ of chance of exiting the only Pb shielding. Above 3 MeV, this percentage is almost constant at $\sim$2.5$\%$ up to 10 MeV. With Pb and BP shielding this percentage is even reduced below 2$\%$ for all energies of incident $\gamma$-rays. The slight dip in the distribution is due to the interplay of competing processes of Compton scattering and pair production for $\gamma$-ray energy loss mechanism in the medium at these energies.
\begin{figure}[h]
\hspace{-0.2in}
\begin{center}
\includegraphics[scale=0.38]{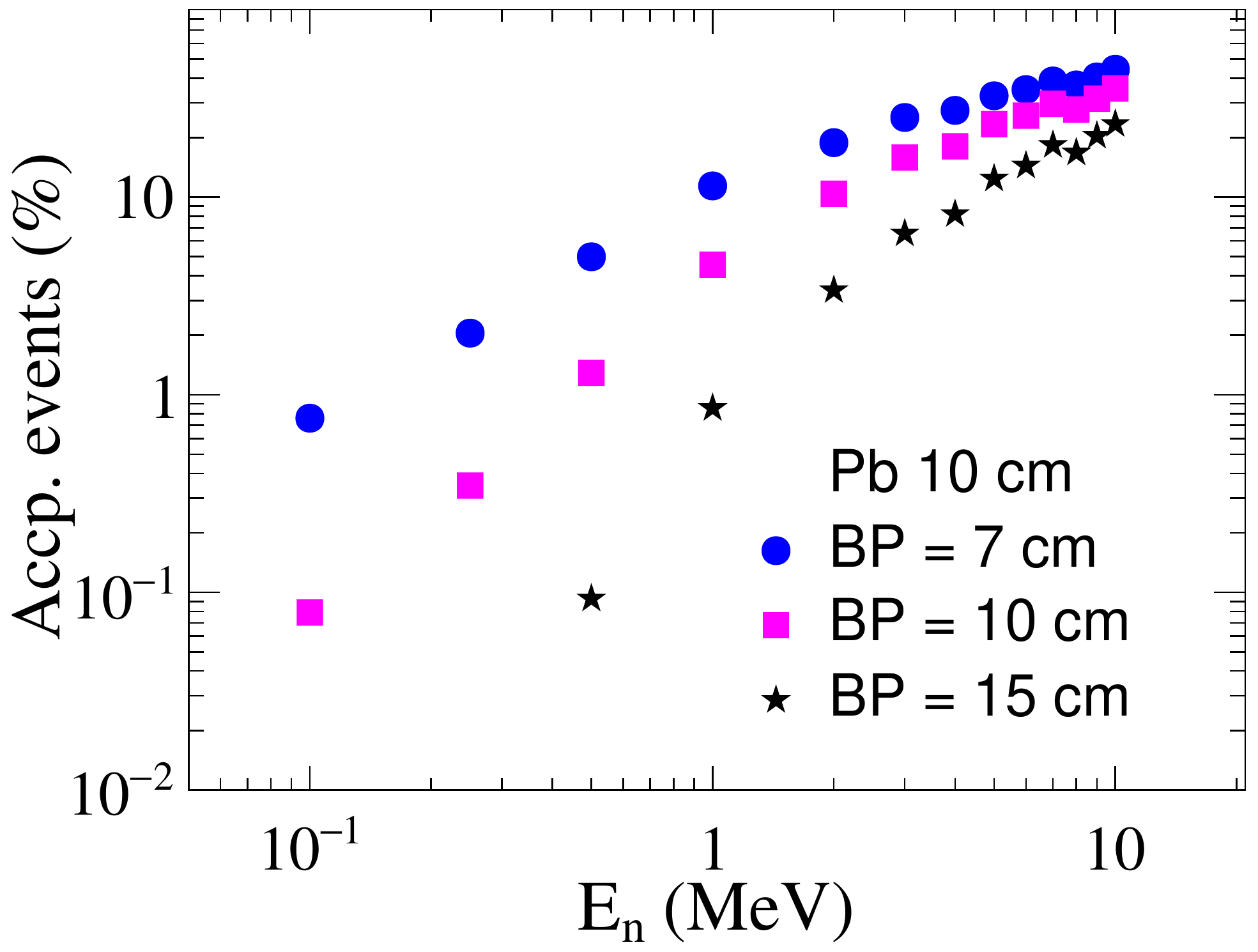}
\caption{Accepted neutrons events, inside ISMRAN detector, for a 10 cm Pb with varying thickness of BP shielding configurations.}
\label{AccepNeut}
\end{center}
\end{figure}
A similar study is also performed for the neutrons using a 10cm Pb and with various thickness of BP shielding in ISMRAN. It can be seen from Fig.~\ref{AccepNeut}, that the low energy neutrons are significantly suppressed, with the increasing BP shield thickness, from entering into the ISMRAN detector volume. Above 1 MeV incident neutron energies, the percentage of the accepted neutron events entering in ISMRAN after passing the BP shielding gradually increases and is almost at 10$\%$. These fast neutrons can be reduced by incorporating an extra shielding of paraffin or high density plastic material. Efforts are underway to incorporate additional shielding, while keeping in view the constraints on the amount of shielding weight allowed by the average load bearing capacity of the floor at the experimental site.   
\section{Pulse processing electronics and data acquisition system}
\begin{figure}[h]
\begin{center}
\includegraphics[scale=0.38]{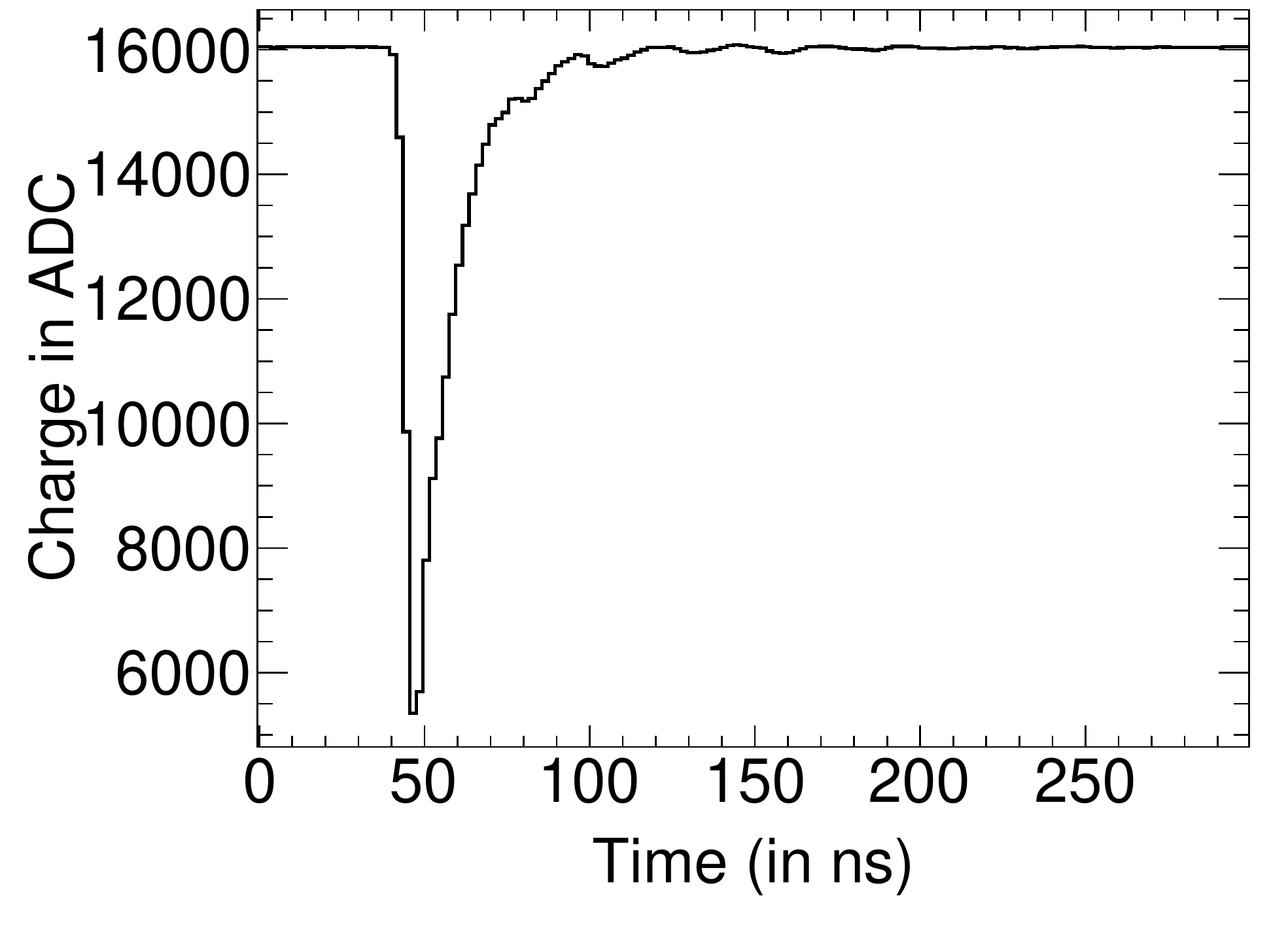}
\caption{Signal waveform from CAEN V1730 digitizer for a PS bar.}
\label{waveform}
\end{center}
\end{figure}

\begin{figure}[h]
\begin{center}
\includegraphics[scale=0.38]{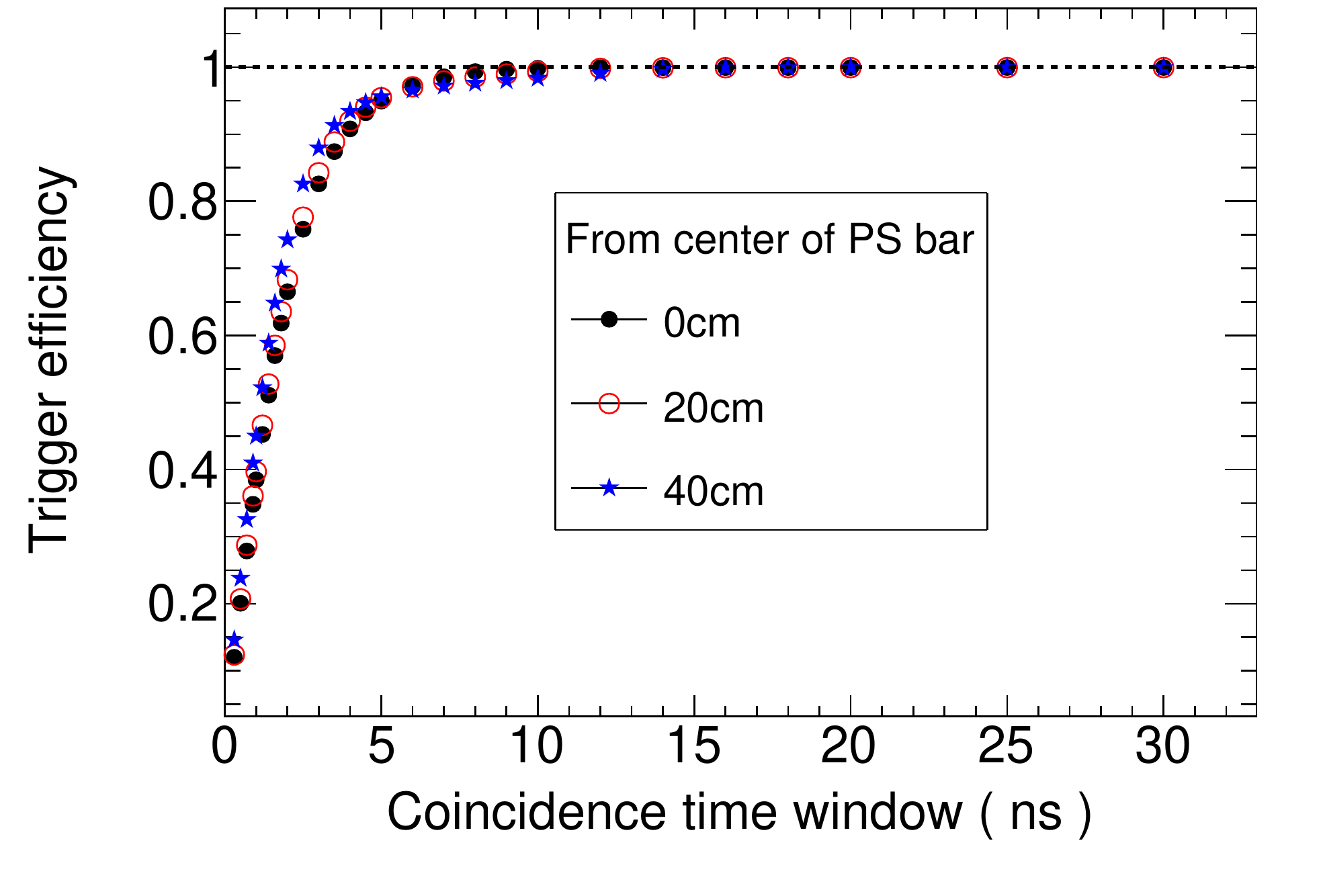}
\caption{Trigger efficiency as a function of coincidence time window between signals from two PMTs of a single PS bar.}
\label{TrigWindow}
\end{center}
\end{figure}

Due to the large number of readout channels for the signals from ISMRAN array, CAEN V1730 16 channel 500 MS/s frequency VME based waveform digitizers are used for pulse processing and data acquisition system. In this system, the discrimination, gate generation, charge integration, timing and coincidence of the PMT signals for each channel are processed by FPGAs. Optical communication and digitizing the waveforms on the board itself allows the data acquisition to be performed at $\sim$ 100MHz rate while the cumulative event rates with shielding are only $\sim$50kHz. The signals from PMTs at both ends of a single bar are read in coincidence on two individual channels of the digitizer. For the mini-ISMRAN, two digitizers, synchronized for timestamps, are used for the DAQ. 
Figure~\ref{waveform} shows a typical waveform obtained from one PMT signal of the PS bar. The charge integration gate range for obtaining the signal in ADC units is chosen to be 150 ns spanning the complete signal duration. Every event recorded, above operating threshold, by each PMT of a single PS bar are timestamped. This allows us to build coincidence events and reduce the random event rates by appropriately selecting the coincidence window width in the digitizer firmware while online filtering of the data from the digitizer boards.  
In order to decide upon the optimum time window for triggering on coincidence events from two PMTs of a single PS bar, the trigger efficiency from data using a ${}^{137}$Cs source is calculated as a function of coincidence time window. The data are recorded for a large coincidence time window of 200 ns in three configurations by keeping the ${}^{137}$Cs source at the center of the bar (0 $\mathrm{cm}$) and at different position towards one end of the bar. Figure~\ref{TrigWindow} shows the trigger efficiency obtained by taking the ratio of number of events triggered in bins of time window from 0.1 to 200 ns with the total number of events recorded. It is found that the trigger efficiency reaches 100$\%$ for coincidence time window of 16 ns or more and is almost independent of the position of the events in the single PS bar for this value onwards. 
The digitizers are programmed with the latest CAEN DPP PSD firmware version $4.11.139.6$ which has been benchmarked using standard scintillators~\cite{RAWAT2016186}. This firmware allows for discrimination of different radiations such as $\gamma$-rays, neutrons or alpha particles if their pulse profiles in the given scintillator volume are different. Due to the use of PS bars in ISMRAN experiment that have almost identical pulse profile for both $\gamma$-rays and neutrons, the PSD technique is not effective. But nonetheless, it is helpful in quantifying the reactor background with liquid scintillator detector where the pulse shape discrimination is possible.

\section{Characterization of plastic scintillator bars}
\begin{figure}
\begin{center}
\includegraphics[scale=0.75]{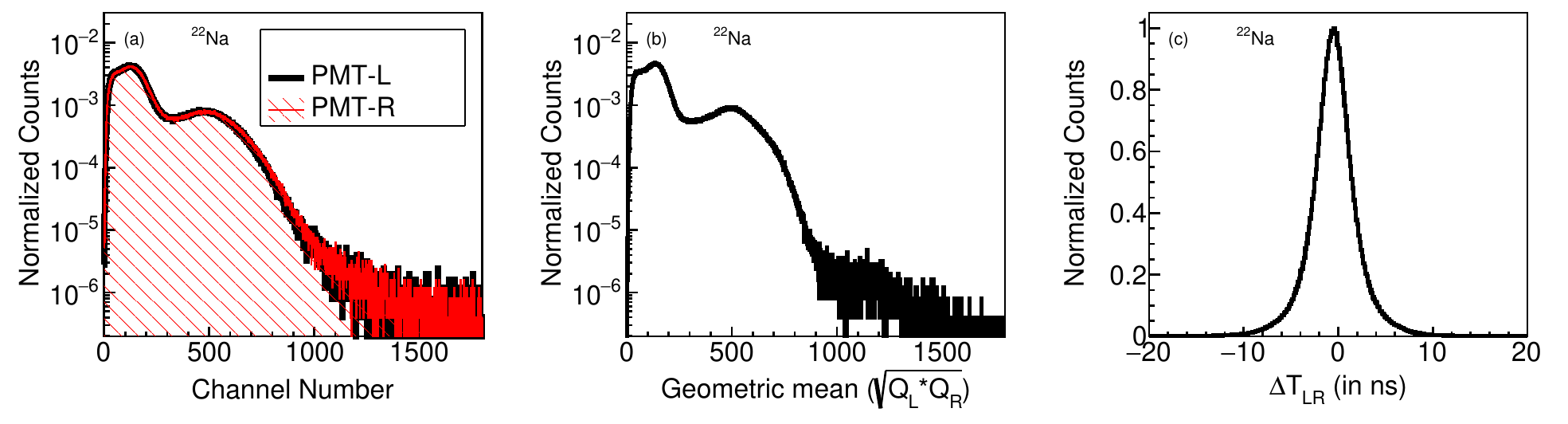}
\caption{(a) ADC distribution from the Left and Right PMTs of a PS bar (b) Geometric mean distribution of ADCs from the two PMT and (c) $\mathrm{\Delta T_{LR}}$ timestamp difference between the two PMT signals of a PS bar. The measurements are done with a ${}^{22}$Na source kept at center of PS bar.}
\label{barprop}
\end{center}
\end{figure}
\begin{figure*}[h]
\begin{center}
\includegraphics[scale=0.78]{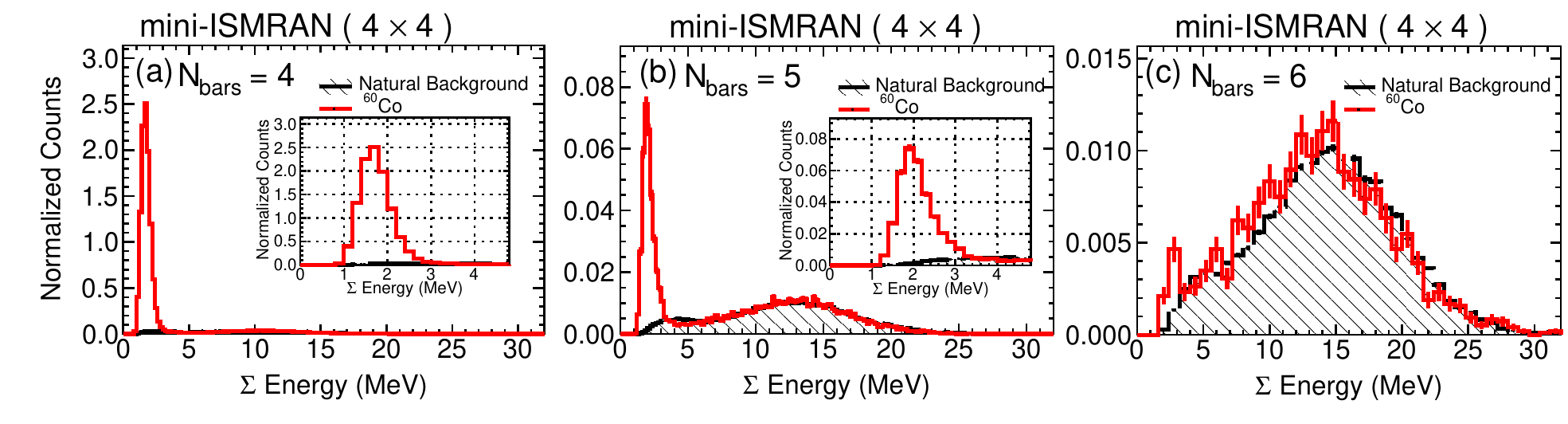}
\caption{The sum energy distribution within 40 $ns$ time window, (a) $\mathrm{N_{bars}}$ = 4, (b) $\mathrm{N_{bars}}$ = 5 and (c) $\mathrm{N_{bars}}$ = 6, for the ${}^{60}\mathrm{Co}$ (solid histogram) and natural background (filled histogram) events. The insets in panel (a) and (b) shows the zoomed x-axis of the sum energy distribution for the ${}^{60}\mathrm{Co}$ source and natural background.}
\label{Co60Sum}
\end{center}
\end{figure*}

In order to understand the energy and timing characteristics of each PS bar across ISMRAN, it is important to do a gain matching along with the energy and time calibrations of individual bars for uniform response across ISMRAN setup. The energy calibration for each PS bar is done using $\gamma$-rays from known radioactive sources such as ${}^{137}$Cs, ${}^{22}$Na and ${}^{60}$Co in the laboratory. The radioactive source is placed on top of the PS bar and the spectra are recorded without any collimation of the emitted $\gamma$-rays. This method introduces additional spread in the timing and the position resolution of the PS bar. Figure~\ref{barprop}(a) and (b) shows the left and right PMTs gain matched ADC spectra in channel number and corresponding geometric mean (GM) distribution taken with ${}^{22}$Na source at the center of a single PS bar, respectively. The gain matching is uniform over the entire range of measured ADC distribution. Clear Compton edges for 0.511 MeV and 1.274 MeV $\gamma$-rays from the ${}^{22}$Na source can be seen. Figure~\ref{barprop} (c) shows the difference in the timestamps ($\mathrm{\Delta T_{LR}}$) from the left and right PMT at each end of a single PS bar when data are recorded in coincidence condition. 
\begin{figure}[h!]
  \begin{center}
\includegraphics[scale=0.38]{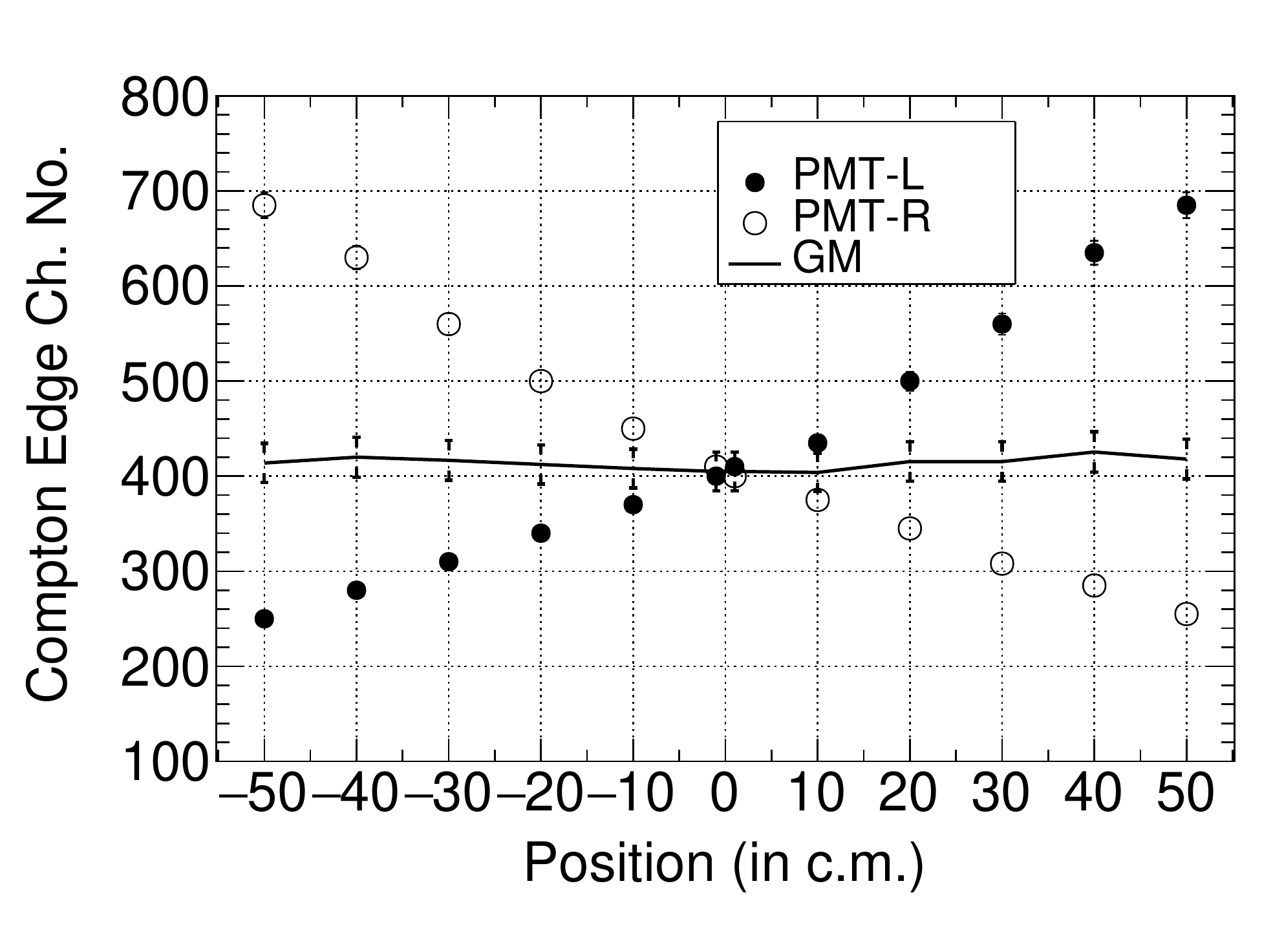}
\caption{Attenuation of the signal in PMTs, left (closed) and right (open), for Compton scattered electron from 0.662 MeV $\gamma$-ray from ${}^{137}$Cs source along 100 cm of a PS bar. The center of the PS bar is taken as 0 cm while the leftmost and rightmost position are marked as +50 cm and -50 cm respectively. Also shown is the GM (solid curve) of the signal from left and right PMTs.}
\label{avgatten}
\end{center}
\end{figure}
\begin{figure*}
  \begin{center}
\includegraphics[scale=0.55]{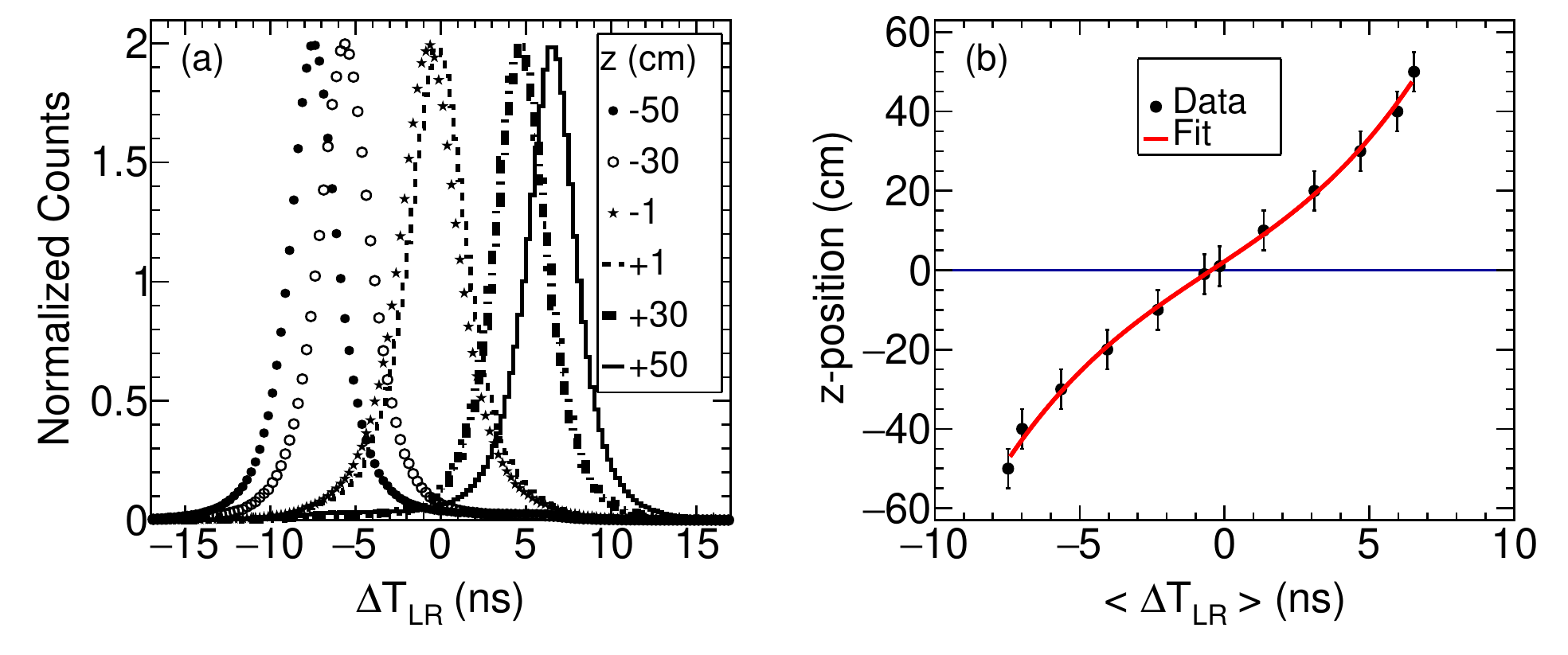}
\caption{(a) $\mathrm{\Delta T_{LR}}$ distribution from left and right PMT timestamps with ${}^{137}$Cs source placed at different positions along the 100 cm length of PS bar. (b) Parametrization of hit position in PS bar as a function of average $\mathrm{\Delta T_{LR}}$.}
\label{zposbar}
\end{center}
\end{figure*}
\begin{figure*}[t]
  \begin{center}
\includegraphics[scale=0.75]{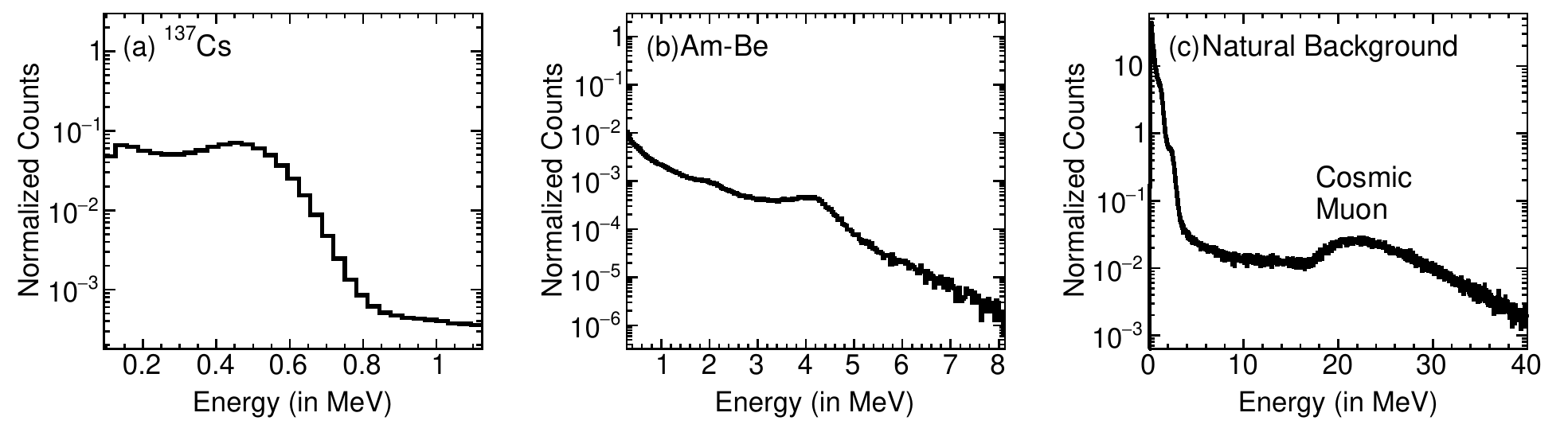}
\caption{Energy distribution of (a) ${}^{137}$Cs, (b) Am-Be, (c) $\gamma$-rays from natural radioactive sources (${}^{40}$K and ${}^{208}$Tl) and minimum ionization energy deposit from cosmic muon in PS bar.}
\label{EnergyBar}
\end{center}
\end{figure*}
The scintillation light produced inside a scintillator, reaches the PMTs at end with a reduced intensity owing to attenuation along its propagation length. This can often be approximated by an exponential decay having an effective attenuation length $\lambda$ specific to the detector material and reflector configuration. However, by taking the geometric mean of the PMT signals from both ends of the PS bar corrects for the effect of light attenuation where optical propagation is well described by exponential function~\cite{KARSCH2001362}. This is particularly effective if $\lambda$ is sufficiently large as compared to the length of the detector, as is the case with EJ-200 scintillator detectors ($\lambda$ $\sim$3.8 m) \cite{eljen}. Figure~\ref{avgatten} shows the average attenuation of the signal in PMTs as a function of L, moving from the middle to one end, of single PS bar. The ADC channel number corresponding to the Compton edge of the single $\gamma$-ray of ${}^{137}$Cs source is plotted as function of source position along the PS bar, which is in steps of 10 cm with center position as 0 cm and leftmost and rightmost positions as +50 cm and -50 cm respectively. As is evident from the plot, the geometric mean, shown by the solid curve, largely corrects for the effect of signal attenuation. Thus the GM of PMT charge signal approximately describes the total charge deposited inside the scintillator for almost 80$\%$ of the central detector volume except at the ends.
With the provision of reading events with timestamps from digitizer for the left and right PMTs of a PS bar, we can obtain the hit position from the parameterization of $\mathrm{\Delta T_{LR}}$ along the length of the bar. Figure~\ref{zposbar} shows the measured average $\mathrm{\Delta T_{LR}}$ at different positions along the bar using a ${}^{137}$Cs radioactive source (in the form of 1 cm wide disk) placed on top of PS bar without collimation. The parameterization is done using a second order polynomial function. With this parameterization and the timing resolution of $\sim$2 ns,  we obtain a position resolution of $\sim$10 cm along the PS bar.
The maximum energy deposited by the Compton scattered electrons are used to identify the structures in the measured charge from the PMTs at the ends of the PS bar. Figure~\ref{EnergyBar} shows the calibrated energy distribution of(a) ${}^{137}$Cs, (b) Am-Be and (c) $\gamma$-rays from natural radioactive sources (${}^{40}$K and ${}^{208}$Tl) of the environmental background. In Fig~\ref{EnergyBar}(b), Compton edge of the 2.2 MeV $\gamma$-ray from the neutron capture on H is seen along with Compton edge of 4.4 MeV $\gamma$-ray at higher energy. In Fig~\ref{EnergyBar} (c), a bump at $\sim$20 MeV is due to the minimum ionization energy deposited by muons in the 10 cm thick PS bar. The average non reactor background rates from different natural sources without shielding are $\sim$27 Hz and $\sim$10 Hz for ${}^{40}$K(1.460 MeV), ${}^{208}$Tl(2.614 MeV) around the Compton edge energies and $\sim$6 Hz for cosmic muons respectively. The integrated non reactor background rate in 3-8 MeV region is $\sim$132 Hz.
\begin{figure}[h!]
\begin{center}
\includegraphics[scale=0.4]{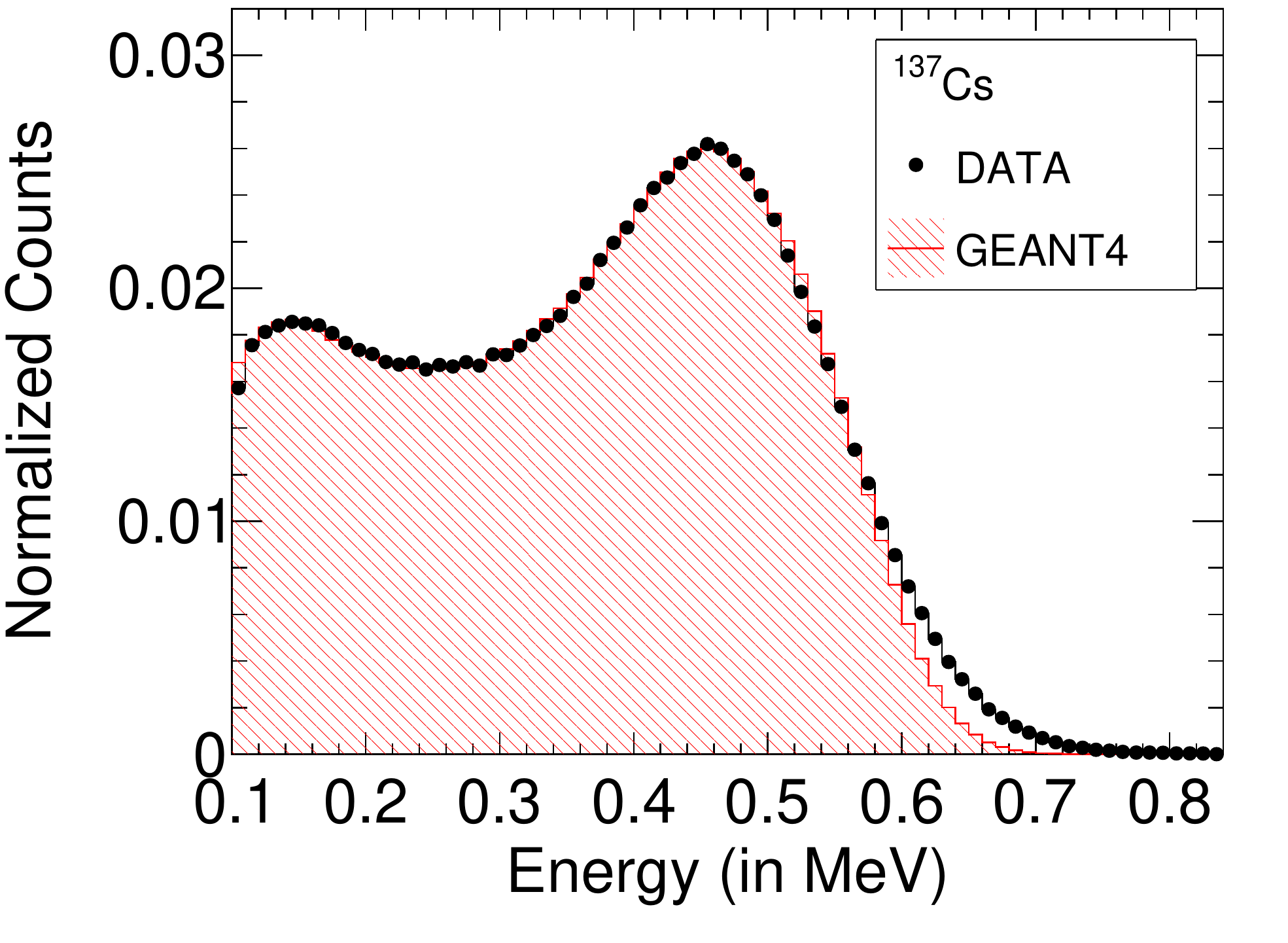}
\caption{Comparison of energy distribution between data (solid) and simulation (dashed) from a ${}^{137}$Cs source spectrum in a 100cm $\times$ 10cm $\times$ 10cm PS bar. The simulation results are smeared for the energy resolution of the plastic scintillator bar.}
\label{DataSimComp}
\end{center}
\end{figure}
In order to validate the simulation results we have compared the energy response of ${}^{137}$Cs between data and simulations. Figure~\ref{DataSimComp} shows the measured calibrated energy distribution from data (solid symbols) and simulated energy distribution (dashed histogram) from GEANT4 simulations for one single PS bar. The simulation results are smeared with a energy resolution function $\sim20\%/\sqrt{E}$ to get a reasonable agreement with the measured data. The disagreement between data and simulation results beyond 0.5 MeV are mainly due to lack of modeling of natural background component in the simulation results. One of requirements of the ISMRAN detector array is to classify prompt and delayed events based on the sum energies in PS bars and $\mathrm{N_{bars}}$ hit. In order to study the feasibilty of reconstructing coincident event, i.e. Compton scattered $\gamma$-ray energy deposition in multiple bars within a time window, in mini-ISMRAN, we use $\mathrm{{}^{60}Co}$ source which emits 1.17 and 1.33 MeV $\gamma$-rays in coincidence. The source is placed at the center of the mini-ISMRAN to maximize the containment of both the coincident $\gamma$-rays from the $\mathrm{{}^{60}Co}$ source. Figure~\ref{Co60Sum}(a), (b) and (c) shows the sum energy distribution in mini-ISMRAN for $\mathrm{{}^{60}Co}$ source and for natural background by requiring $\mathrm{N_{bars}}$ = 4, $\mathrm{N_{bars}}$ = 5 and $\mathrm{N_{bars}}$ = 6 condition within a time window of 40 ns, respectively. Only those PS bars are selected for the sum energy where the individual energy deposit in each PS bar is between 0.2 to 7.5 MeV, to reduce the cosmic muon background. As it can be seen from Fig~\ref{Co60Sum}(a) and Fig~\ref{Co60Sum}(b), a peak at $\sim$2 MeV corresponds to the coincident $\gamma$-ray events from $\mathrm{{}^{60}Co}$ source. The reconstructed sum energy of the coincident event improves with $\mathrm{N_{bars}}$ = 5 PS bars as compared to $\mathrm{N_{bars}}$ = 4 PS bars in the sum energy distribution as shown in zoomed version of the sum energy distribution in the inset of Fig~\ref{Co60Sum}(a) and Fig~\ref{Co60Sum}(b). The inefficiency in sum energy reconstruction is due to energy threshold $\mathrm{E^{Th}_{bar}}$ $>$ 0.2 MeV on individual PS bars and due to the finite acceptance of mini-ISMRAN resulting in reconstruction of mean sum energies to values lower than 2.5 MeV. Figure~\ref{Co60Sum}(c) shows the sum energy distribution for $\mathrm{N_{bars}}$ = 6 condition where it can be seen the fraction of coincident events reconstructed from $\mathrm{{}^{60}Co}$ are reduced significantly as compared to the natural background events. Also it is observed that the ratio of signal from coincident events with $\mathrm{{}^{60}Co}$ source to natural background events worsens with increase in $\mathrm{N_{bars}}$ in sum energy distribution. This study shows, by appropriately selecting  $\mathrm{N_{bars}}$ and energy deposited in each PS bar the coincident events can be reconstructed in mini-ISMRAN. In all the three $\mathrm{N_{bars}}$ condition, the sum energy distribution above 5 MeV from $\mathrm{{}^{60}Co}$ and natural background scales indicating the common source of background events. 
  
\section{Reactor background measurements}
\begin{figure}[h]
\begin{center}
\includegraphics[scale=0.38]{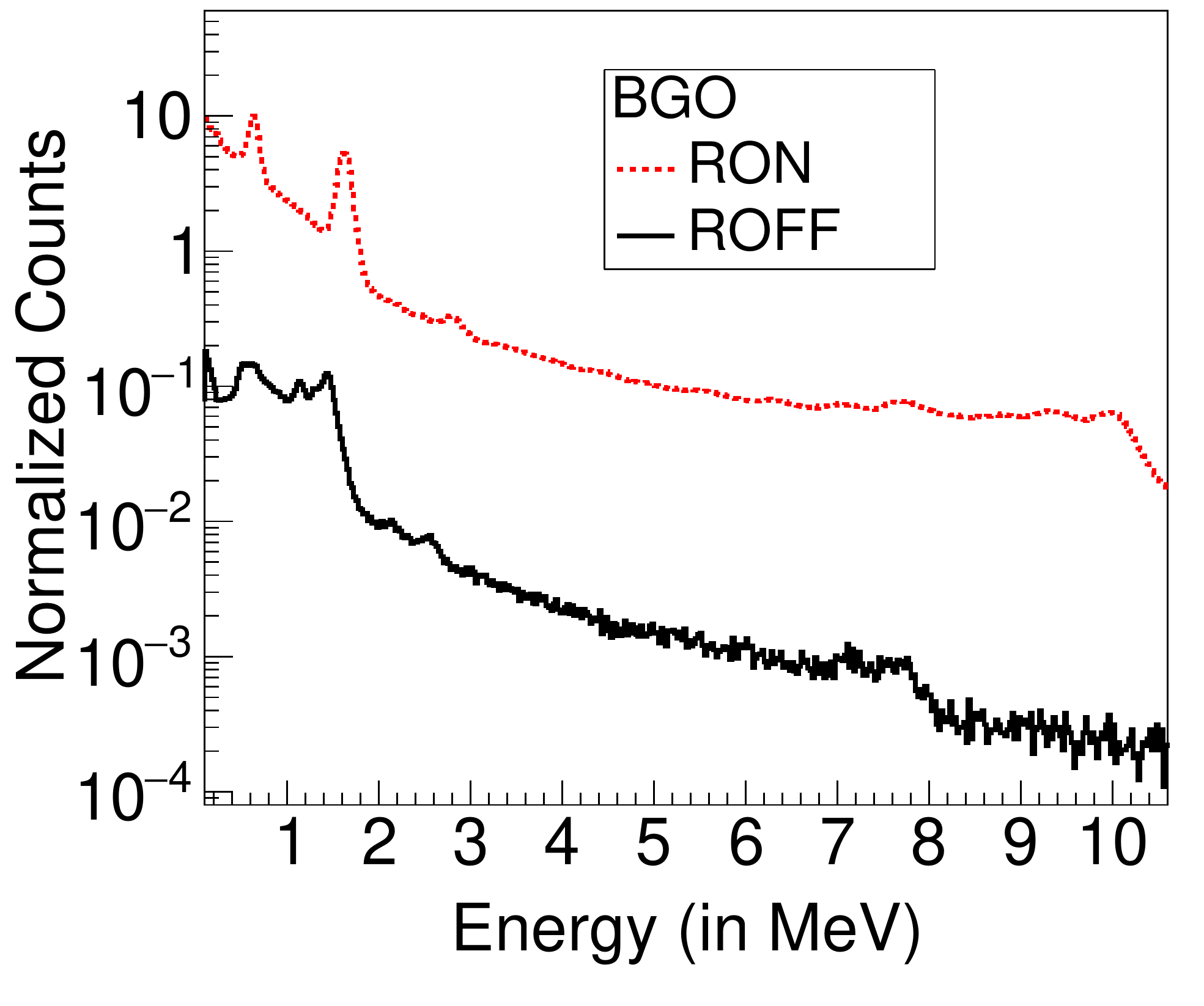}
\caption{Energy distribution from a BGO detector in reactor ON (dashed) and OFF (solid) condition.}
\label{BGOOnOff}
\end{center}
\end{figure}

Reducing the reactor related background is crucial for detection of the $\overline\nuup_{e}$ candidate events. Measurements to quantify the background level in reactor ON and OFF conditions are performed with a Bismuth Germanate (BGO) crystal coupled to a PMT tube and a single PS bar from the mini-ISMRAN array. All the measurements, with PS bar and BGO placed side-by-side, comparing the data between reactor ON and OFF are performed in Pb (10cm) + BP (10cm) enclosed shielding. Present mini-ISMRAN setup is located at $\sim$13m distance from the reactor core. The location of the mini-ISMRAN setup is in proximity to a quasi elastic neutron scattering setup relying on the thermal neutrons from a beam port. The flux of thermal neutrons from the beam port is $\mathrm{\sim 10^{5} n/cm^{2}/sec}$. The scattered neutron paths are well shielded by a beam dump. The effect of background from the beam port is also studied by taking data in reactor ON condition with and without a Cadmium shield on the beam port. The difference in the background rate measured in one PS bar facing the experiment is observed to be negligible in both the conditions.
\begin{figure}[h]
\begin{center}
\includegraphics[scale=0.38]{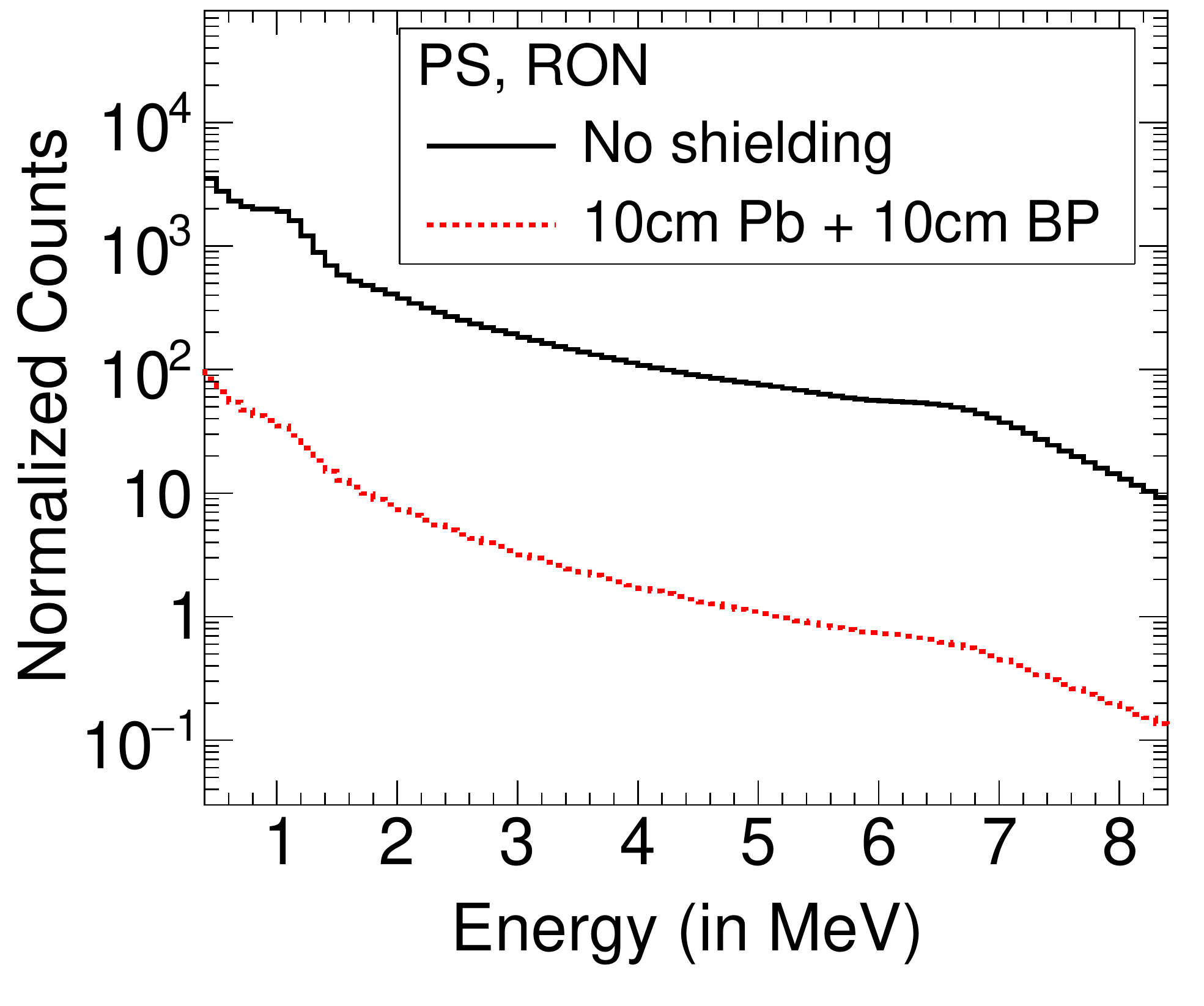}
\caption{Energy distribution from a single PS bar of mini-ISMRAN in reactor ON condition. The measurements are compared between the data from no shielding (solid) and for 10 cm Pb and 10 cm BP (dashed) shielding.}
\label{PSShielding}
\end{center}
\end{figure}
The background measurements for one PS bar of the mini-ISMRAN array are performed with and without 10 cm Pb and 10 cm BP shielding. Figure~\ref{PSShielding} shows the energy distribution from such a PS bar without any shielding in solid histogram, and with 10 cm Pb and 10 cm BP in dashed histogram. The integrated background rate without any shielding is $\sim$24 kHz in the energy range starting from 0.2 MeV up to 40 MeV and with 10 cm Pb and 10 cm BP shielding this rate is reduced to $\sim$500 Hz at full power.
To further quantify the $\gamma$-ray background, the measurements are carried out using a BGO detector in reactor ON and OFF conditions inside a 10 cm Pb and 10 cm BP shielding. The BGO detector is placed next to the PS bar in the shielded mini-ISMRAN setup. Figure~\ref{BGOOnOff} shows the energy distribution of measured $\gamma$-ray background from BGO detector in reactor ON and OFF condition. Above 3 MeV, there are small structures in the energy distribution in reactor ON conditions which are absent in reactor OFF condition. The increase in the $\gamma$-ray background in the reactor ON condition is attributed to the $\gamma$-ray coming from the thermal neutron capture on the surrounding material in the reactor and experimental hall. A detailed study of the  $\gamma$-ray energy spectrum from neutron capture on different materials in the reactor hall can be found in reference~\cite{HFIR}. Some of these $\gamma$-ray lines can be seen in our measurements from the BGO detector in reactor ON condition.
\begin{figure}[h]
\begin{center}
\includegraphics[scale=0.38]{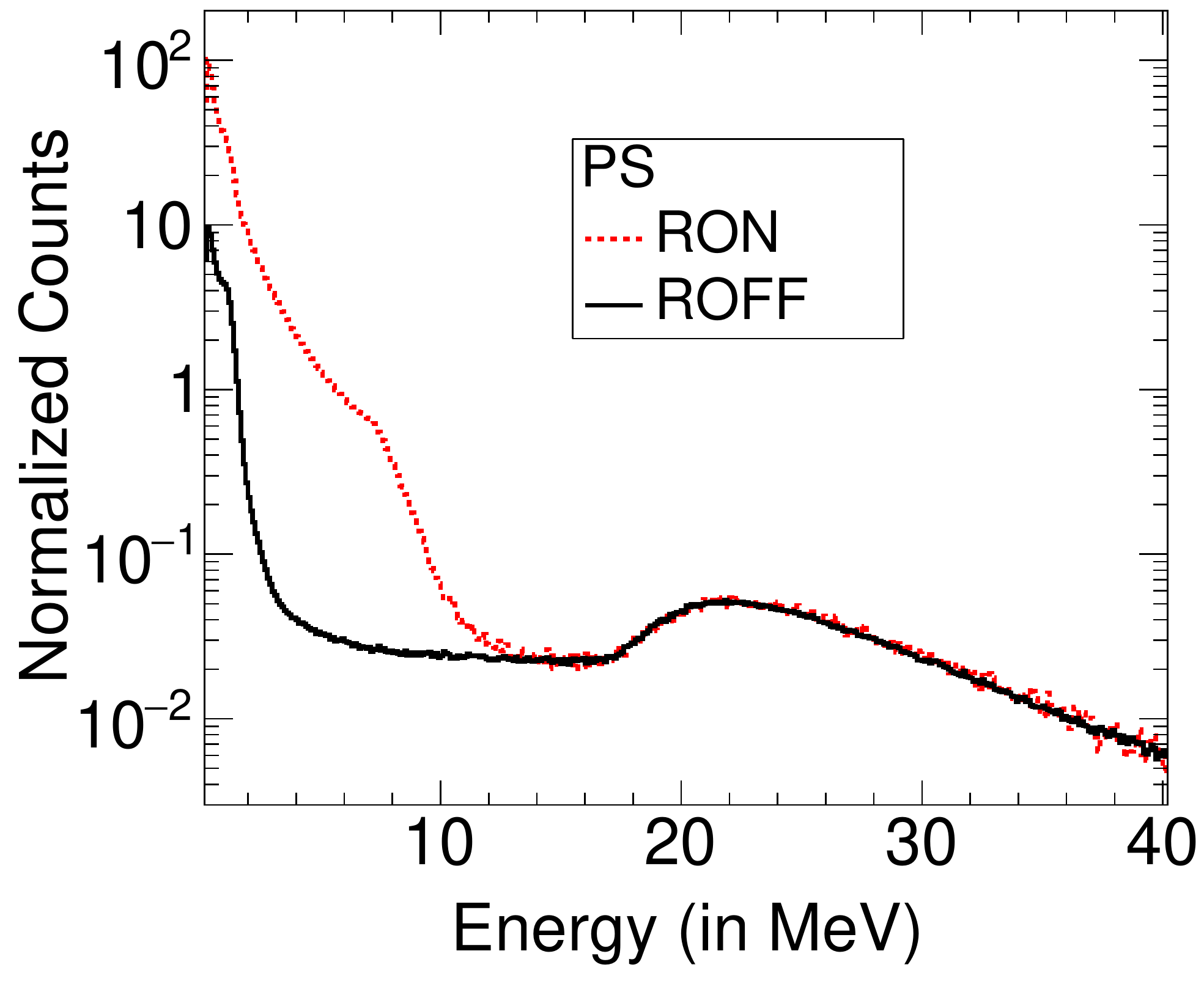}
\caption{Energy distribution from a PS bar in reactor ON (dashed) and OFF (solid) condition.}
\label{PSOnOff}
\end{center}
\end{figure}

\begin{table}[h]
  \begin{center}
  \caption{Background rates measured in mini-ISMRAN for various shielding configurations in reactor ON condition.}
  \label{Table2}
  \renewcommand{\arraystretch}{0.75}
\begin{tabular}{ |p{5.5cm}|p{1.5cm}| }
\hline
Detector configurations & Count Rates (Hz) \\
\hline
No Shielding (Single PS bar) & $\sim$ 24,000 \\
\hline
10 cm thick lead shield & $\sim$ 2,000 \\
\hline
10 cm thick lead + 10 cm thick B.P. & $\sim$ 500 \\
\hline
10 cm thick lead + 10 cm thick B.P. ($\mathrm{N_{bars}}$ = 2, time window $<$ 40 ns) &$\sim$ 10 \\
\hline
\end{tabular}
\end{center}
\end{table}

Figure~\ref{PSOnOff} shows the energy distribution of background with a single PS bar from mini-ISMRAN array inside the 10 cm Pb and 10 cm BP shielding for reactor ON and OFF conditions in wider energy range. In general, the background is more (factor of 10) in  reactor ON condition as compared to reactor OFF data. Most of the reactor ON background at higher energies up to 7 MeV comes from $\gamma$-rays emanating from the neutron capture events in the surrounding material in the reactor hall as explained above for the $\gamma$-ray spectrum from BGO detector in reactor ON condition. Due to poorer energy resolution as compared to BGO, the distinct peaks are not visible in the PS bar energy spectrum in reactor ON condition. Above 10 MeV, there is very slight difference in the measured background with PS bar in reactor ON and OFF condition, indicating that the dominant source in this region is from cosmic muons and purely non-reactor related background. The measurement of background rates done with different configurations of shielding for single PS bar and event selection criteria on two PS bars from the center of the mini-ISMRAN array are shown in table~\ref{Table2}. The reactor related background reduces from kHz to 500 Hz with shielding of 10cm Pb and 10cm of BP. To reduce background rate further a condition on $\mathrm{N_{bars}}$ = 2 hit within a time window $<$ 40 ns is used for the event selection. As from the simulation studies it can be seen, that positron signal is mostly contained in two bars, the above event selection criteria provides an estimate of random rate in selecting prompt-like events in reactor ON condition. With this mentioned event selection criteria the background rate is reduced to $\sim$10Hz. 

\begin{figure}
\begin{center}
\includegraphics[scale=0.75]{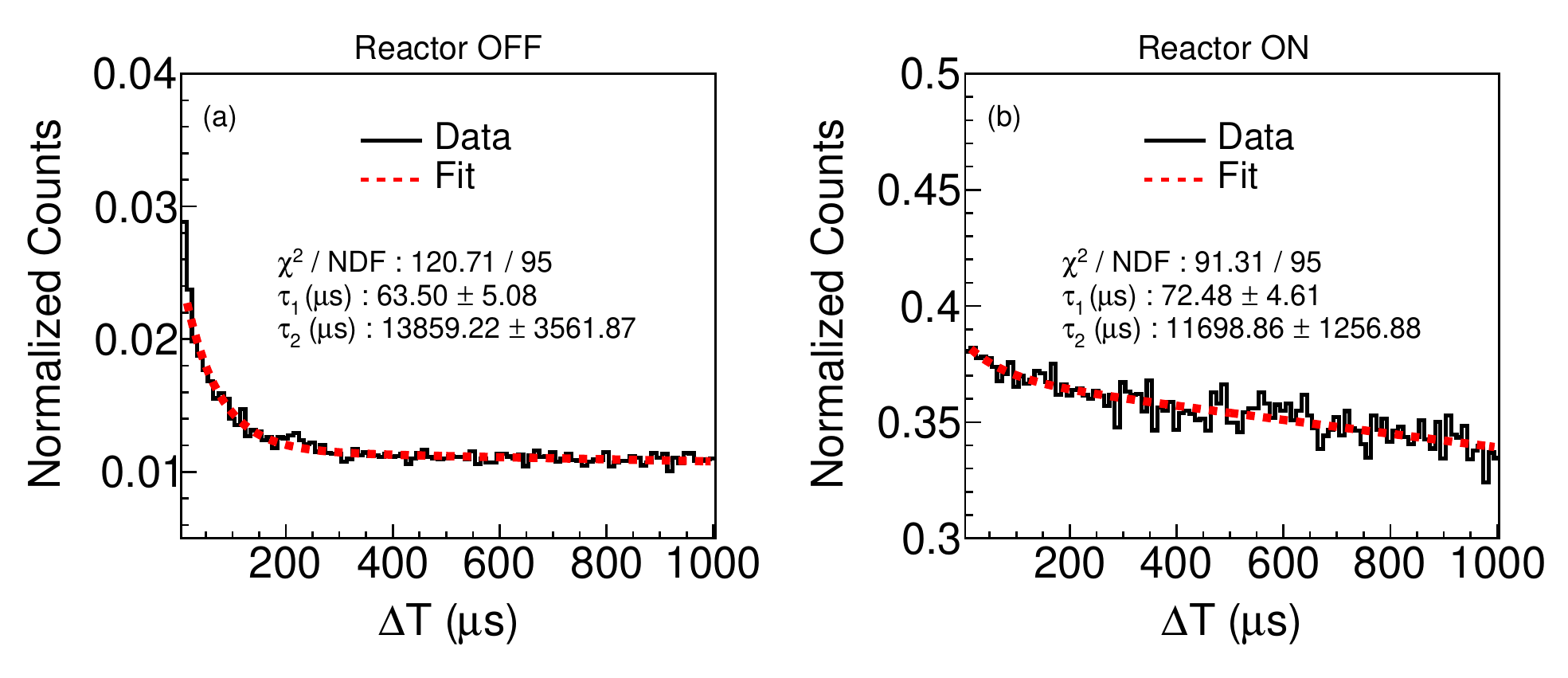}
\caption{$\mathrm{\Delta T}$ time difference between two PS bars under (a) reactor OFF and (b) reactor ON condition. The $\mathrm{\Delta T}$ distributions are fitted with two exponential (dashed) function.}
\label{DelTROnROff}
\end{center}
\end{figure}
Apart from sum energy and $\mathrm{N_{bars}}$ based event selection criteria to identify $\overline\nuup_{e}$ candidate events, time difference between prompt and delayed events can also be used for distinguishing the correlated and uncorrelated component. The time correlation distribution can have different components arising from different background processes. Short time constant is dominated by the cosmogenic background involving fast neutrons or from neutron spallation from the surrounding material in the reactor hall. The prompt event time constant can have the contamination from the fast neutron recoil, accidental neutron capture or from $\gamma$-ray de-excitation of a nucleus excited by a fast neutron inelastic scattering process. The delayed event time constant are primarily from neutron capture events in the detector volume. A very large time constant indicates the uncorrelated background in the detector. To understand the time correlation in reactor environment, we have performed measurement of the timing differences between two adjacent PS bars in the center of mini-ISMRAN array, in reactor ON and OFF conditions. The two selected PS bars, with energy between 0.2 MeV to 7.5 MeV, would act like proxy for prompt and delayed event signatures. The selection on energy of individual PS bar is applied to get spectral uniformity in energy and to reduce contamination from cosmic muons in the event selection. Prompt and delayed events are selected with in a time window of 4.0 $<$ $\mathrm{\Delta T}$ ($\mu s$) $<$ 1000. Figure~\ref{DelTROnROff}(a) and (b) shows the timing difference ($\mathrm{\Delta T}$) distribution between the two adjacent PS bars from the mini-ISMRAN array in ROFF and RON condition, respectively. Both the distributions are time normalized. The large difference in normalized counts and relative suppression in the correlated component at short time constant between RON and ROFF condition is due to increase in singles background rate for PS bars in RON as compared to ROFF condition. The RON time correlation distribution is mostly dominated by background. Further study on event selection criteria with sum energy and $\mathrm{N_{bars}}$ variable needs to be done in RON condition. The $\mathrm{\Delta T}$ distributions are fitted with a two exponential function with $\tau_{1}$ and $\tau_{2}$ representing the short and long time constants, respectively. The $\mathrm{\Delta T}$ distributions are fitted in region from 10 $\mu$s to 1000 $\mu$s to avoid background from muon decay events in ISMRAN. It is found that $\tau_{1}$, the short component, in reactor OFF and ON condition between two bars is 63.50$\pm$5.08$\mu$s and 72.48$\pm$4.61, respectively. A long time constant $\tau_{2}$ in both cases, reactor OFF and ON, indicates the random background in RON and OFF condition.
\section{Conclusions and Outlook}
A 1.0 ton detector, Indian Scintillator Matrix for Reactor Anti-Neutrino (ISMRAN), is being developed to measure reactor $\overline\nuup_{e}$ at the Dhruva research reactor facility at BARC. The core detector consists of 10$\times$10 array of Gd-wrapped plastic scintillator bars. This experiment will demonstrate the capability of above ground remote monitoring of reactors and is an effort towards setting guidelines for development of such systems at power reactors. Detailed simulations for the events from positron annihilation as well as neutron capture in ISMRAN experiment are presented in terms of energy deposited in PS bars and the number of bars hit above threshold. Simulation studies to optimize shielding thickness of lead and borated polyethylene are carried out with ISMRAN setup. From initial measurements at the reactor site, it is shown that with 10cm of Pb and 10 cm BP, a sizable fraction of reactor and non reactor related background are reduced significantly. The use of waveform digitizers for data acquisition system is implemented for ISMRAN setup to maximize the event rate handling capacities in real time experiment. The basic component of the detector, plastic scintillator bars coupled with PMTs on both ends, is characterized to study the energy response, timing and position sensitivity of the signal in the ISMRAN setup. A prototype detector, mini-ISMRAN, is used for measurement of non-reactor based background in laboratory and reactor related backgrounds at the Dhruva reactor hall. With $\mathrm{{}^{60}Co}$ source, the event reconstruction of two coincident $\gamma$-rays in mini-ISMRAN in non reactor environment is done by selection on deposited energy and number of PS bars in a time window of 40ns. For the comparison of $\gamma$-ray backgrounds at reactor hall, measurements are made with BGO and plastic scintillator detectors in reactor ON and OFF conditions. Time correlation measurements between events in two PS bars for reactor ON and OFF condition have been carried out for estimation of short and long time constants. The full ISMRAN setup is expected to be commissioned soon and will be in data taking mode by end of the year.
\section{Acknowledgments}
We are thankful to Research Reactor Services Division, BARC for logistic support and co-operation, Centre for Design and Manufacture, BARC for taking up design and fabrication of the final ISMRAN support structure. We are also thankful to V. M. Datar, Tata Institute of Fundamental Research, A. K. Mohanty, Saha Institute of Nuclear Physics, A. Bernstein and N. Bowden from Lawrence Livermore National Laboratory for useful discussions and suggestions on development of ISMRAN setup.
\section{References}
\bibliographystyle{unsrt}  

\end{document}